\documentclass[sigconf, 9pt]{acmart}
\copyrightyear{2023}
\acmYear{2023}
\setcopyright{acmlicensed}
\acmConference[CCS '23] {Proceedings of the 2023 ACM SIGSAC Conference on Computer and Communications Security}{November 26--30, 2023}{Copenhagen, Denmark.}
\acmBooktitle{Proceedings of the 2023 ACM SIGSAC Conference on Computer and Communications Security (CCS '23), November 26--30, 2023, Copenhagen, Denmark}
\acmPrice{15.00}
\acmISBN{979-8-4007-0050-7/23/11}
\acmDOI{10.1145/3576915.3623089}
\renewcommand\footnotetextcopyrightpermission[1]{} 
\usepackage{amsmath}

\usepackage{amssymb,amsfonts}
\usepackage{algorithmic}
\usepackage{graphicx}
\usepackage{textcomp}
\usepackage{xcolor}
\def\BibTeX{{\rm B\kern-.05em{\sc i\kern-.025em b}\kern-.08em
    T\kern-.1667em\lower.7ex\hbox{E}\kern-.125emX}}
\usepackage[utf8]{inputenc}
\usepackage{booktabs}
\usepackage{graphicx}
\usepackage{subfigure}
\usepackage{units}
\usepackage{accents}
\usepackage{blindtext}
\usepackage{multirow}
\usepackage{xcolor,xspace}
\usepackage{amsmath}
\usepackage{paralist}
\usepackage{mdframed}
\usepackage{pict2e}
\usepackage{listings}
\usepackage{color}
\usepackage{threeparttable}
\usepackage{array}
\usepackage{booktabs}
\usepackage{pifont}
\usepackage{balance}
\usepackage{url}
\usepackage{tabularx}
\usepackage{caption}
\usepackage{lipsum}
\usepackage{makecell}
\usepackage[most]{tcolorbox}
\usepackage{amsbsy}

\usepackage[symbol]{footmisc}
\renewcommand{\thefootnote}{\fnsymbol{footnote}}

\newcommand\blfootnote[1]{%
  \begingroup
  \renewcommand\thefootnote{}\footnote{#1}%
  \addtocounter{footnote}{-1}%
  \endgroup
}

\usepackage[ruled]{algorithm2e}
\usepackage{algorithmic}
\LinesNumbered

\usepackage{url}

\usepackage[
    n,
    advantage,
    operators,
    sets,
    adversary,
    landau,
    probability,
    notions,
    logic,
    ff,
    mm,
    primitives,
    events,
    complexity,
    asymptotics,
    keys]{cryptocode}

\usetikzlibrary{arrows, patterns, automata}

\definecolor{burgundy}{RGB}{144,0,32}
\definecolor{myblue}{rgb}{0.0, 0.2, 0.8}
\definecolor{bluegray}{rgb}{0.4, 0.6, 0.8}
\definecolor{azure}{rgb}{0.0, 0.5, 1.0}
\definecolor{darkcandyapplered}{rgb}{0.64, 0.0, 0.0}

\newcommand{\paisa}{{{\sf PAISA}}\xspace}

\newcommand{\dev}{\ensuremath{dev}\xspace}
\newcommand{\svr}{\ensuremath{svr}\xspace}

\newcommand{\iotdev}{\ensuremath{I_{\dev}}\xspace}
\newcommand{\iotdevpk}{\ensuremath{pk_{\iotdev}\xspace}}
\newcommand{\iotdevsk}{\ensuremath{sk_{\iotdev}\xspace}}
\newcommand{\usrdev}{\ensuremath{U_{\dev}}\xspace}
\newcommand{\mfr}{\ensuremath{M_{\svr}}\xspace}
\newcommand{\mfrpk}{\ensuremath{pk_{\mfr}\xspace}}
\newcommand{\mfrsk}{\ensuremath{sk_{\mfr}\xspace}}
\newcommand{\iotdevsoft}{\ensuremath{SW_{\dev}\xspace}}
\newcommand{\iotdevid}{\ensuremath{ID_{\dev}\xspace}}
\newcommand{\iotdevsofthash}{\ensuremath{H_{\iotdevsoft}\xspace}}
\newcommand{\paisatcb}{\ensuremath{SW_{\paisa}\xspace}}

\newcommand{\devmanifest}{\ensuremath{\mathsf{Manifest_{\iotdev}}}\xspace}
\newcommand{\devmanifestsig}{\ensuremath{\mathsf{Sig_{Man}}}\xspace}
\newcommand{\devmanifesturl}{\ensuremath{\mathsf{URL_{Man}}}\xspace}
\newcommand{\devmanifesturlfull}{\ensuremath{\mathsf{URL_{Man_{Full}}}}\xspace}
\newcommand{\hashfunc}{\ensuremath{{\ensuremath{\sf{\mathcal H}}}\xspace}}
\newcommand{\sigfunc}{\ensuremath{{\ensuremath{\sf{\mathcal SIG}}}\xspace}}
\newcommand{\sigvalue}[1]{\ensuremath{\mathsf{Sig_{#1}}}\xspace}
\newcommand{\noncevalue}[2]{\ensuremath{\mathsf{N_{#1}^{#2}}}\xspace}
\newcommand{\tsvalue}[1]{\ensuremath{\mathsf{time_{#1}}}\xspace}

\newcommand{\syncreq}{{\ensuremath{\sf{\mathcal SyncReq}}}\xspace}
\newcommand{\syncresp}{{\ensuremath{\sf{\mathcal SyncResp}}}\xspace}
\newcommand{\syncack}{{\ensuremath{\sf{\mathcal SyncAck}}}\xspace}
\newcommand{\announce}{{\ensuremath{\sf{\mathcal Announce}}}\xspace}
\newcommand{\anno}{\ensuremath{anno}\xspace}
\newcommand{\announceinterval}{{\ensuremath{\sf{T_{\announce}}}}\xspace}
\newcommand{\announcemessage}{{\ensuremath{\sf{Msg_{\anno}}}}\xspace}
\newcommand{\attest}{{\ensuremath{\sf{\mathcal Attest}}}\xspace}
\newcommand{\attestresult}{\ensuremath{\mathsf{Att_{result}}}\xspace}
\newcommand{\attestreport}{\ensuremath{\mathsf{Att_{report}}}\xspace}
\newcommand{\attestinterval}{{\ensuremath{\sf{T_{\attest}}}}\xspace}

\newcommand{\verifymessage}{{\ensuremath{\sf{\mathcal Verify}}}\xspace}

\newcommand{\registration}{{\ensuremath{\mathsf{\it Registration}}}\xspace}
\newcommand{\boot}{{\ensuremath{\mathsf{\it Boot Time}}}\xspace}
\newcommand{\runtime}{{\ensuremath{\mathsf{\it Runtime}}}\xspace}

\newcommand{\provision}{\ensuremath{\mathsf{\bf Provision}}\xspace}
\newcommand{\timesync}{\ensuremath{\mathsf{\bf Time Sync}}\xspace}
\newcommand{\announcement}{\ensuremath{\mathsf{\bf Announcement}}\xspace}
\newcommand{\reception}{\ensuremath{\mathsf{\bf Reception}}\xspace}
\newcommand{\initdev}{\ensuremath{\mathsf{\bf Init Device}}\xspace}

\newcommand{\prv}{{\ensuremath{\sf{\mathcal Prv}}}\xspace}
\newcommand{\vrf}{{\ensuremath{\sf{\mathcal Vrf}}}\xspace}
\newcommand{\ra}{{\ensuremath{\sf{\mathcal RA}}}\xspace}
\newcommand{\sadv}{{\ensuremath{\sf{\mathcal Adv}}}\xspace}
\newcommand{\chal}{{\ensuremath{\sf{\mathcal Chal}}}\xspace}

\newcommand{\privacyaware}{{\it privacy-agile}}

\mathchardef\mhyphen="2D

\newtheorem{protocol}{Protocol}

\begin{document}

\title{Caveat (IoT) Emptor\footnotemark{}: \\ Towards Transparency of IoT Device Presence (Full Version)}

\author{Sashidhar Jakkamsetti}
\email{sashidhar.jakkamsetti@us.bosch.com}
\affiliation{%
   \institution{University of California, Irvine\footnotemark{}}
}

\author{Youngil Kim}
\email{youngik2@uci.edu}
\affiliation{%
   \institution{University of California, Irvine}
}

\author{Gene Tsudik}
\email{gene.tsudik@uci.edu}
\affiliation{%
   \institution{University of California, Irvine}
}

\begin{abstract}
As many types of IoT devices worm their way into numerous settings and many aspects of our daily lives,
awareness of their presence and functionality becomes a source of major concern.
Hidden IoT devices can snoop (via sensing) on nearby unsuspecting users, and impact
the environment where unaware users are present, via actuation. This prompts, respectively,
privacy and security/safety issues. The dangers of hidden IoT devices have been recognized
and prior research suggested some means of mitigation, mostly based on traffic analysis or 
using specialized hardware to uncover devices. While such approaches are partially effective, 
there is currently no comprehensive approach to IoT device transparency. 

Prompted in part by recent privacy regulations (GDPR and CCPA), this paper motivates and constructs a \privacyaware\  Root-of-Trust architecture
for IoT devices, called {\bf \paisa}: \underline{P}rivacy-\underline{a}gile \underline{I}oT \underline{S}ensing and \underline{A}ctuation. It guarantees timely and 
secure announcements about IoT devices' presence and their capabilities.
\paisa has two components: one on the IoT device that guarantees periodic 
announcements of its presence even if all device software is compromised, 
and the other that runs on the user device, which captures and processes announcements.
Notably, \paisa requires no hardware modifications; it uses a popular 
off-the-shelf Trusted Execution Environment (TEE) -- ARM TrustZone.
This work also comprises a fully functional (open-sourced) prototype implementation of 
\paisa, which includes: an IoT device that makes announcements via 
IEEE 802.11 WiFi beacons and an Android smartphone-based app
that captures and processes announcements. Both security and performance of
\paisa design and prototype are discussed. 
\blfootnote{$^{*}$ `Caveat Emptor'' is Latin for "User Beware".}
\blfootnote{$^{\dagger}$ Currently at Robert Bosch LLC -  Research and Technology Center.}

\end{abstract}
\renewcommand{\thefootnote}{\arabic{footnote}}

\maketitle

\section{Introduction}\label{sec:intro}
Internet of Things (IoT) and embedded (aka "smart") devices have become an integral part of modern 
society and are often (and increasingly) encountered in many spheres of everyday life, including homes,
offices, vehicles, public spaces, ports, and warehouses. It is estimated that, by 2030, 
there will be over 29 billion Internet-connected IoT devices \cite{statista-iot}.

Unlike general-purpose computers, IoT devices are specialized and their main functions involve
some forms of sensing and/or actuation.  Some of them perform 
safety-critical tasks and collect sensitive personal information.
IoT device manufacturers understandably prioritize (novel) functionality, external aesthetics, 
ease-of-use, and other factors, while security is usually treated as a secondary issue or an afterthought. 
This is partly due to various constraints, including physical space, energy, and monetary cost. 

Unsurprisingly, IoT devices represent attractive attack targets, 
e.g., \cite{attacks1,stuxnet,mirai,attacks2,cisco}. Attacks vary widely and generally aim to 
compromise security/safety, privacy, or just to zombify targeted devices, e.g., as in the Mirai botnet.
In recent years, a lot of research effort has been invested into mitigating security and privacy 
issues in the IoT ecosystem. Many proposed techniques 
(e.g., \cite{authiot2023,iftls,jin2017virtual,smartcamprivacy,neto2016aot,privacyiothealth}) 
use end-to-end encryption, authentication, and other cryptographic constructs 
to secure IoT devices.
Another research direction focused on protecting sensitive data from passive in-network adversaries, 
(e.g. \cite{trimananda2020packet,apthorpe2018keeping,apthorpe2017closing}) 
by performing analysis based on traffic metadata.
There is also a large body of research on mitigating software compromise of devices: 
remote attestation (e.g., \cite{tan2011tpm,smart,pioneer,pistis,simple}), run-time integrity attestation 
(e.g., \cite{oat,geden2019hardware,tinycfa,litehax,cflat,lofat}), and sensor data protection (e.g., \cite{sancus,pfb}).

All of the above are merely research proposals. Although
device manufacturers sometimes integrate research-originated techniques 
into their products, they rarely acknowledge the adoption of external research results.  
Furthermore, there are no strong compelling factors nudging the manufacturers towards
adoption of security features.

Although there are several guidelines\footnote{For example, NIST Recommended 
Criteria for Cybersecurity Labeling for Consumer IoT Products 
\cite{nist-reco} and UK/Australia Code of Practice for Consumer IoT Security 
\cite{aus-code-practice,uk-code-practice}.} for IoT security, 
they do not consider user privacy in the general sense. Such well-intentioned guidelines are aimed 
at device owners or operators, who are generally well aware of device placement and
capabilities. However, IoT devices impact {\bf all human users} in their vicinity by 
sensing them and/or controlling their environment.

This occurs in public places, such as parks, public transport, office buildings, concert halls, stadiums, and airports. 
It also happens in less-public places, such as hotels and private rentals, e.g., Airbnb. 
In the latter, users tend to be wary of unfamiliar surroundings \cite{inman,enduser-privacy} 
partly because they are unaware of nearby devices, their capabilities, what data exactly is being 
collected, and how it is (or will be) used. In particular, the issue of undeclared and hidden
cameras has plagued the private rental industry \cite{iotprivacyperception}.

We believe that, ideally, there would be an agreed-upon means of informing nearby (and thus 
potentially impacted) users about the presence of IoT devices as well as their capabilities
and current activities. This would facilitate an informed decision by the users, i.e., 
whether to stay or leave the IoT-instrumented space.

\subsection{Motivation}
Based on the preceding discussion, the main motivation for this work is the need to take 
a step towards a privacy-compliant IoT ecosystem where all impacted users are made
aware of nearby IoT devices, which empowers them to make informed decisions.
Another inspiration stems from recent data protection regulations, such as the 
European General Data Protection Regulation (GDPR) \cite{gdpr} and California Consumer Privacy 
Act (CCPA) \cite{ccpa}. 
These regulations aim to protect user privacy by 
stipulating that service providers must be accountable and ask for user consent before collecting, 
processing, storing, and sharing user data.  We want to apply the same principle to IoT devices.

Note that these regulations are clearly focused on privacy, meaning that, in the IoT context, 
they naturally apply to devices that sense the environment. Whereas, our scope is broader --
it includes actuation-capable devices that can directly impact nearby users' security and even safety.
For example, consider a situation where a hotel guest with epilepsy is unaware of a ``smart'' fire/smoke alarm in the room
which turns on a strobe light when it detects smoke or fire. Unexpected light strobing can easily
cause an epileptic seizure or worse.\footnote{Ideally, the guest who is warned about the alarm could switch
it to another mode, without dire consequences.} Another example is an Airbnb renter who is unaware of
a smart door-lock that can be (un)locked remotely which presents a risk of the door being closed or opened
without the renter's knowledge. Whereas, if forewarned, the renter could disable it for 
the period of stay. To this point, a 2017 incident with an Austrian Hotel where all smart locks
were hacked illustrates the danger.\footnote{See: \url{https://www.bbc.com/news/business-42352326}}

Addressing privacy concerns in the IoT context poses two challenges:
\begin{compactenum}
\item How to make users aware of the presence of nearby devices? 
\item How to ask for consent to: collect information (in case of sensing), or control the environment 
(in case of actuation)?
\end{compactenum}
In this paper, we take the first step by focusing on (1), while viewing (2) as its natural follow-up.  
Current means of achieving (2) mostly focus on obtaining user 
consent~\cite{hong2004architecture,pa2,iotprivacy-ppa,iotprivacy-designspace}. 
For example, studies on 
Privacy Assistants\cite{hong2004architecture,pa2,iotprivacy-ppa} focus on automating the process of 
acquiring user preferences/consent efficiently. 
Another research direction\cite{habib2021toggles,utz2019informed,iotprivacy-designspace} provides design (and implementation) guidelines for user privacy choices that address regulatory considerations.

Regarding (1), there are several approaches for informing users about ambient devices.
One approach involves manually scanning the environment using specialized hardware\cite{bugdetector,NLJD,mmwave,camdetector}. 
Another way is by monitoring wireless traffic, i.e., 
WiFi and/or Bluetooth \cite{lumos,snoopdog,cameraspy}. Though somewhat effective, such techniques 
are cumbersome and error-prone, since it is not always possible to thoroughly scan the entire
ambient space. Also, these approaches can be evaded if a device is mis-configured or compromised.
Nevertheless, they represent the only option for discovering hidden and non-compliant devices.

Instead of putting the burden on the users to monitor and analyze wireless traffic, we want to
construct a technique that guarantees that all compliant IoT devices reliably announce their presence, which 
includes their types and capabilities. Consequently, a user entering 
an unfamiliar space can be quickly warned about nearby IoT activity. We believe that this is an
important initial step towards making future IoT devices privacy-compliant. 
We imagine later integrating the proposed technique with other consent-seeking platforms.

\subsection{Overview \& Contributions}
We construct a technique called \paisa: {\bf P}rivacy-{\bf A}gile {\bf I}oT 
{\bf S}ensing and {\bf A}ctuation, that guarantees timely and secure announcements 
about IoT device presence and capabilities. We use the term {\bf \privacyaware} to 
denote \paisa service -- explicit user awareness of all nearby \paisa-compliant IoT devices. 
Each \paisa-compliant device reliably broadcasts secure announcements at regular intervals, ensuring continuous 
awareness, unless it is compromised via physical attacks or is powered off.

\paisa has two main components: (1) one on the IoT device that guarantees periodic 
announcements of its presence, and (2) the other that runs on the user device (smartphone); 
it captures and processes announcements. To guarantee secure periodic announcements on the IoT device, 
\paisa relies on the presence of a 
Trusted Execution Environments (TEE) or some other active Root-of-Trust (RoT) component. 
The TEE ensures guaranteed and isolated execution of \paisa Trusted Computing Base (TCB). 
On the user device, \paisa imposes no special requirements to capture and process  
announcements: it simply uses standard network drivers to read announcement packets 
and validate them in an application.

Anticipated contributions are:
\begin{compactitem}
    \item Motivation for, and comprehensive treatment of, a \privacyaware\ RoT architecture for IoT devices. 
  To the best of our (current) knowledge, no prior work systematically approached 
  privacy compliance in the IoT ecosystem, given that relevant attempts \cite{lumos,snoopdog,devicemien2019,cameraspy}, 
  are either ad-hoc or not applicable to a wide range of devices.
  \item Design and construction of \paisa, a secure and \privacyaware\ TEE-based architecture that reliably
  informs nearby users about IoT devices. Notably, \paisa does not require any custom hardware, unlike
  some prior work, e.g., \cite{pfb,garota}. It uses {\it off-the-shelf} \ popular 
  TEE, e.g., ARM TrustZone \cite{ARM-TrustZone-M}. 
  \item A fully functional prototype implementation of \paisa, which includes: (a) a prototype IoT device 
  based on ARM Cortex-M33 featuring announcements via IEEE 802.11 WiFi beacons, and (b) an Andriod application 
  running on Google Pixel 6, which extracts and displays the announcements 
  to the user. All source code is publicly available at \cite{paisa-code}.
\end{compactitem}

\subsection{Scope, Limitations, \& Caveats}
As with most new designs, \paisa has certain limitations:
\begin{compactitem}
\item With regard to scope, it applies to a class of devices equipped with some basic security features, e.g., 
ARM TrustZone. Thus, it is unsuitable for simple ``bare-metal'' devices or even slightly higher-end ones
that lack a secure hardware element. 
\item In terms of the security level, it offers protection against hacked (directly re-programmed) or malware-infected
devices. However, it does not defend against non-compliant devices. This includes devices that are home-made, 
jerry-rigged, or produced by non-compliant manufacturers. 
\item Furthermore, \paisa does not defend against local jamming or {\em wormhole} attacks 
\cite{wormhole1,wormhole4}.\footnote{A Wormhole attack occurs when an announcement from one device is tunneled 
into a remote network and re-announced there, making it appear that the device is present.}
The latter is nearly impossible to thwart. 
However, we propose a method to partially handle these attacks in Sections \ref{subsec:adversary} and \ref{subsec:protocol}.
\item Finally, we do not explore policy issues and implications, i.e., the focus is on  reliably informing
users about adjacent devices. What users do with that information is left to future work.
While we acknowledge that a practical system must include this component, space limitations make it hard
to treat this topic with the attention it deserves.
\end{compactitem}

\section{Background}\label{sec:background}
\subsection{Targeted IoT Devices}\label{subsec:background_iotdevs}

\begin{figure}[t]
  \centering
  \captionsetup{justification=centering}
  \includegraphics[width=1\columnwidth]{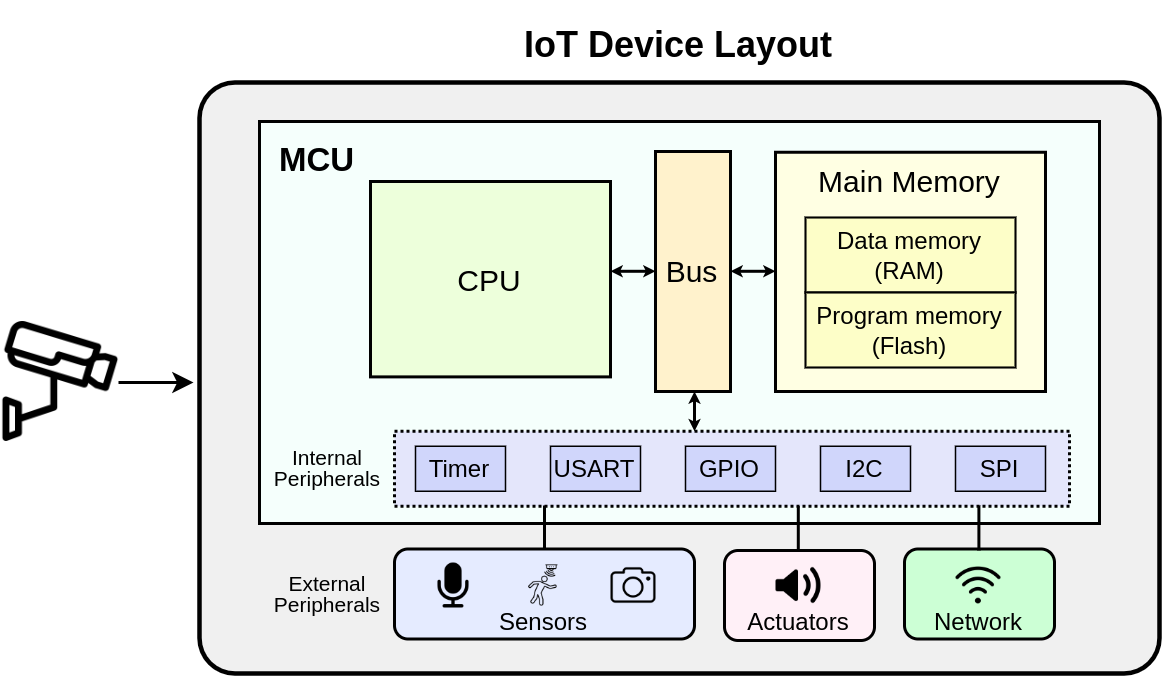}
  \vspace{-0.3cm}
  \caption{\small Architecture of an IoT Device. This example shows the peripherals of a security camera.}
  \label{fig:architecture-device}
  \vspace{-0.4cm}
\end{figure}

This work focuses on resource-limited IoT devices that have strict cost and energy constraints.
Such devices tend to be deployed on a large scale and are meant to perform simple tasks, e.g., thermostats, 
security cameras, and smoke detectors. Due to the constraints, they are often equipped with micro-controller 
units (MCU), such as ARM Cortex-M series \cite{cortexM}. Nonetheless, our work is also applicable to higher-end computing 
devices (e.g., smartwatches, drones, and infotainment units) that are equipped with a TEE.
Recall that very simple devices that have no security features are out of scope.

Figure \ref{fig:architecture-device} shows a general architecture of a device with an MCU and 
multiple peripherals. An MCU is a low-power computing unit that integrates a core processor, 
main memory, and memory bus on a single System-on-a-Chip (SoC). Its main memory is usually
divided between program memory (or flash) where the software resides, and data memory (or RAM), 
which the software uses for its stack, heap, and peripheral memory access. 
A typical MCU also contains several internal peripherals such as a timer, General-Purpose 
Input/Output (GPIO), Universal Asynchronous Receiver/Transmitter (UART), 
Inter-Integrated Circuit (I2C), and Serial Peripheral Interface (SPI).

\noindent {\bf Sensors \& Actuators:}
Multiple purpose-specific sensors and actuators are connected to the MCU via internal peripherals.
While sensors collect information from the environment, actuators control it. 
Examples of sensors are microphones, GPS units, cameras, as well as smoke and motion detectors. 
Examples of actuators are speakers, light switches, door locks, alarms, and sprinklers.

\noindent {\bf Network Interfaces:}
IoT devices are often connected to the Internet and other devices, either directly or via a controller hub or a router. 
Thus, they are typically equipped with at least one network interface (such as WiFi, Bluetooth, Cellular, Ethernet, or Zigbee) attached to the MCU via internal network peripherals, e.g., UART, I2C, or SPI.
WiFi and Cellular are used for wireless Internet connectivity at relatively high speeds.
Bluetooth and Zigbee are used for relatively low-speed short-range communication with other devices,
e.g., a smartphone for Bluetooth, or a controller hub for Zigbee. 
Since WiFi is currently the most common interface available for IoT devices\cite{blevswifi}, \paisa uses 
it for broadcasting device announcements. 
However, any other broadcast media (wired or wireless) can be supported; see Section \ref{sec:limitation}
for more details.

\begin{table*}[h]
  \vspace{-.7em}
  \centering
  \footnotesize  
  \resizebox{\textwidth}{!}{ 
     \begin{tabular}{|c||c|c|c|} \hline
     {\bf IoT device} & {\bf Sensor} & {\bf Actuator} & {\bf Network I/F}    \\ 
     \Xhline{2\arrayrulewidth} 
     X-Sense smart smoke detector \cite{xsense-smoke-detector} & smoke, carbon monoxide detector & alarm & WiFi \\ \hline
     Amazon smart plug \cite{amazon-plug} & - & switch & WiFi \\ \hline
     Blink Mini Security Camera \cite{blink-security-camera} & microphone, motion, camera & speaker & WiFi \\ \hline
     Google Nest thermostat \cite{google-nest-thermostat} & light, motion, temperature, humidity & heating, cooling & WiFi \\ \hline
     iRobot Roomba 694\cite{irobot-roomba-694} & cliff, dirt, optical & brush/vaccum motor, drive motor & WiFi \\ \hline
     Fitbit - fitness tracker \cite{fitbit-fitness-tracker} & accelerometer, heart rate monitor, GPS, altimeter & vibrating motor, speaker & Bluetooth \\ \hline
     Wyze Lock Bolt - smart lock \cite{wyze-lock-bolt} & fingerprint & lock, speaker & Bluetooth \\ \hline
     \end{tabular}
  }
  \caption{\small Various Types of IoT Devices with different Sensors, Actuators, and Network Interface.} 
  \label{table:iotdevs}
  \vspace{-1.5em}
\end{table*}

Table \ref{table:iotdevs} shows some examples of (low-end) commodity IoT devices with sensors, actuators, and their network interfaces.

\subsection{Trusted Execution Environments (TEEs)}\label{subsec:background_trusted}
A TEE is a hardware-enforced primitive that protects the confidentiality and integrity of sensitive software and 
data from untrusted software, including user programs and the OS. 
Similar to some prior work \cite{tz-regulating,tz-sara,tz-m-sbi,cflat}, we use ARM TrustZone-M as the TEE for the
\paisa prototype. TrustZone-M is available on ARM Cortex-M23/M33/M55 MCUs\cite{ARM-TrustZone-M}. 
However, any TEE that offers trusted peripheral interfaces can be used instead. 

\noindent {\bf ARM TrustZone-M}\label{subsec:background_tz}
ARM TrustZone partitions the hardware and software within the MCU into 
two separate isolated regions: Secure and Normal. The former contains trusted security-critical code and data, 
while the latter houses user programs (or the device software).
The MCU switches between secure and non-secure modes when accessing Secure and Normal regions, respectively. 
TrustZone hardware controllers prevent the MCU from accessing memory assigned to Secure region when it is 
running in non-secure mode, resulting in a secure execution environment. Moreover, at boot time, TrustZone verifies the integrity of trusted code via secure boot and always begins executing from the Secure region 
before jumping into the Normal region. TrustZone for ARMv8-M MCUs is called TrustZone-M (TZ-M).

TZ-M features non-secure callable functions (NSC) for Normal region software to invoke trusted code.
Also, TZ-M can lock internal peripherals into the Secure region making them inaccessible to the Normal 
region via the TrustZone Security Controller (TZSC) that, when configured at boot, maps desired peripherals 
into the Secure region.  This mapping configuration is controlled by TZSC and is checked by the 
secure-boot process at boot time. Furthermore, interrupts attached to secure peripherals are always 
directed to the corresponding Interrupt Service Routines (ISR) in the Secure region.
Also, TrustZone Illegal Access Controller (TZAC) raises a SecureFault exception, when a security violation 
is observed, to the Nested Vectored Interrupt Controller (NVIC) which is then securely processed by 
exception handlers.

\paisa relies on TZ-M for enabling a secure execution environment for its TCB and for implementing secure peripherals.
For a comprehensive overview of TrustZone, see \cite{TZ-overview}. 

\noindent {\bf Other Active Roots-of-Trust (RoTs)}
Active RoTs prevent security violations, unlike their passive counterparts that detect them
\cite{vrasedp,geden2019hardware,oat,seshadri2004swatt}. TEEs are considered active RoTs since 
they prevent violations by raising hardware-faults/exceptions, which are handled in the 
secure mode. Besides TEEs, some active RoTs have been proposed in the research literature, 
e.g.,\cite{sancus,garota,awdt-dominance,casu}. Notably, GAROTA\cite{garota} and AWDT\cite{awdt-dominance} 
offer guaranteed execution of secure ISRs when a configured peripheral is triggered.
Although the current focus is on off-the-shelf devices,
we believe that \paisa can be applied to either GAROTA or AWDT devices. 
Section \ref{sec:limitation} discusses the applicability of \paisa to other architectures. 

\subsection{Remote Attestation (\ra)}\label{subsec:remote_attestation}
\ra is a security service that enables the detection of malware presence on 
a remote device (\prv) by allowing a trusted verifier (\vrf) to remotely measure software running on \prv. 
\ra is a challenge-response protocol, usually realized as follows: 
\begin{compactenum}  
  \item \vrf sends an \ra request with a challenge (\chal) to \prv.
  \item \prv receives the attestation request, computes an authenticated integrity check over its software memory region (in program memory) and \chal, and returns the result to \vrf.
  \item \vrf verifies the result and decides if \prv is in a valid state.
\end{compactenum}
The integrity check is performed by computing either a Message Authentication Code (e.g., HMAC) or a 
digital signature (e.g., ECDSA) over \prv's program memory. Computing a MAC requires \prv to share a 
symmetric key with \vrf, while computing a signature requires \prv to have a private key with the 
corresponding public key known to \vrf. Both approaches require secure key storage on \prv.
\ra architectures for low-end MCUs\cite{sancus,vrasedp} use MACs whereas higher-end TEEs (e.g.,  
Intel SGX\cite{sgx} and AMD SEV\cite{sev}) use signatures.

\paisa uses \ra to ensure integrity of normal device operation, i.e. the device software 
controlling sensors and actuators. However, \paisa relies on TZ-M on the MCU to perform attestation locally, 
instead of via an interactive protocol. Also, it uses signatures to report the attestation result, 
similar to \cite{sgx,sev}.

\section{Design Overview} \label{sec:overview}
\begin{figure}[t]
  \centering
  \captionsetup{justification=centering}
  \includegraphics[width=1\columnwidth]{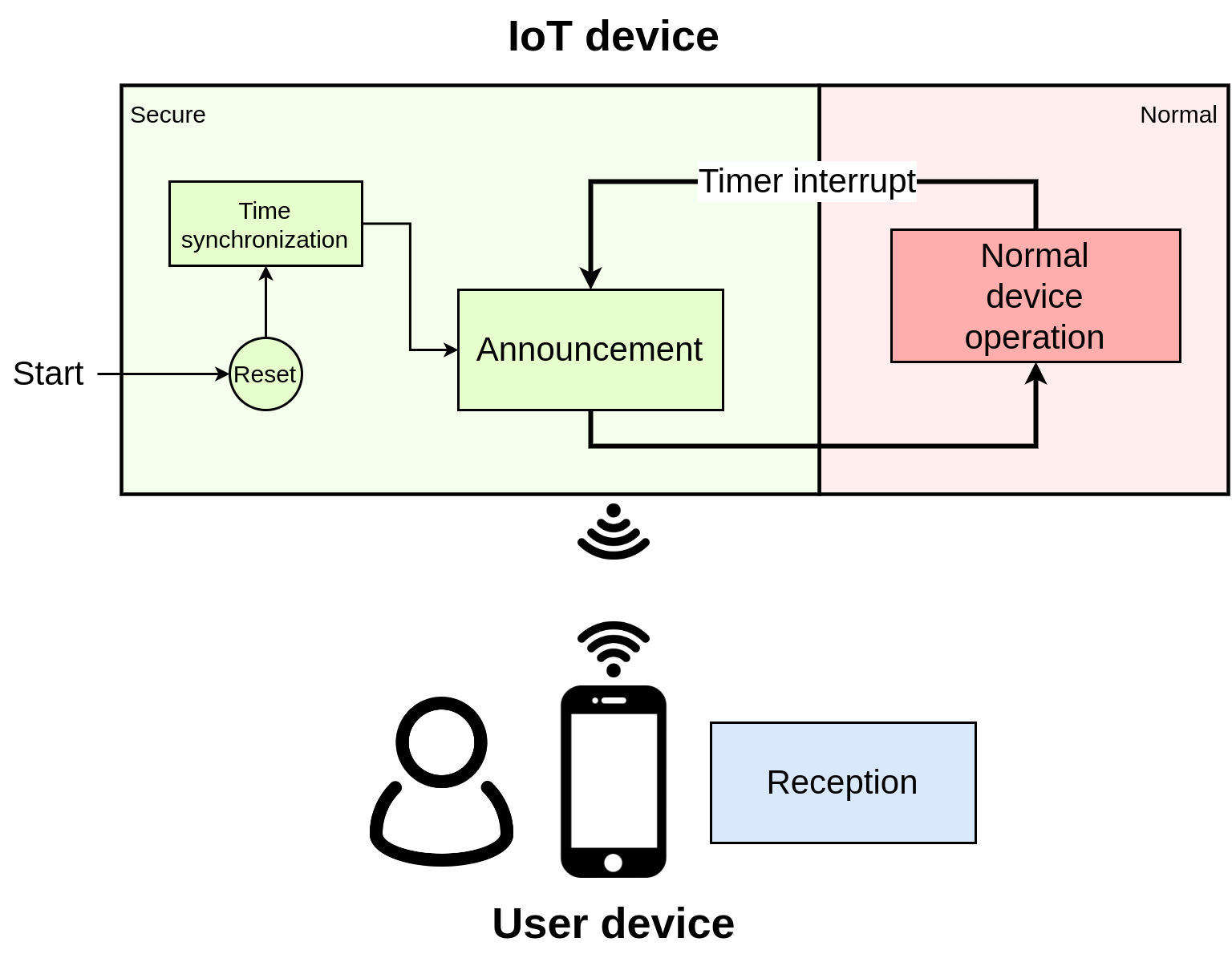}
  \vspace{-.3cm}
  \caption{\small Overview of \paisa workflow.}
  \label{fig:iotdev-flow}
  \vspace{-.3cm}
\end{figure}

\paisa primarily involves two parties: an IoT device (\iotdev) and a user device (\usrdev), e.g., 
a smartphone or a smartwatch.
\paisa is composed of two modules: {\it announcement} on \iotdev and {\it reception} on \usrdev. \\

\noindent {\it \underline{Announcement}}: 
On \iotdev, the {\it announcement} module is trusted and housed inside a TEE.
It ensures that, at periodic intervals, \iotdev broadcasts an announcement 
to other devices within its immediate network reach. Such "reach", i.e. distance, is specified by 
the network interface, e.g., 802.11 WiFi beacons go up to 100 meters \cite{wifibeacon}.
Importantly, \paisa guarantees that announcement packets are broadcast in a timely manner, even if all
device software is compromised. This is achieved via a secure timer and a secure network interface, 
available on TZ-M. 

An announcement packet consists of a fresh timestamp, a device description (sensors, actuators, and their purpose) 
and a signature that authenticates the origin of the packet as a legitimate \iotdev. \\

\noindent {\it \underline{Reception}}: 
On \usrdev, the {\it reception} module captures the announcement packet via its network interface (of the 
same type as on \iotdev). The module then parses the packet, validates its timestamp and signature, and 
conveys the presence of \iotdev and functionality to the user. \\

The proposed design presents some challenges:

\noindent {\bf Device State \& Attestation:}
Merely broadcasting static information, such as a device description, is not enough. If \iotdev software is compromised, 
information disseminated via announcement packets is invalid since \iotdev software does not match
the device description. For example, consider a user who enters an Airbnb rental and learns about 
a motion detector/tracker from \paisa announcements. Suppose that this motion detector is compromised and 
the malware notifies the adversary about the user's presence and movements. 
To handle such cases, the user needs authentic real-time information about the software
running on \iotdev at the announcement time.  Therefore, \paisa attests \iotdev software and includes 
the timestamped attestation report in the announcement. 
The reception module on \usrdev must check the attestation report as part of validating the announcement. 
If the attestation check fails, \iotdev must be compromised and cannot be trusted,
regardless of the description in the announcement.

\noindent {\bf Replay Attacks \& Freshness:}  
To protect against replay attacks and establish freshness of announcements (via timestamps), \iotdev needs 
a reliable source of time. However, a real-time clock is generally not viable for resource-constrained devices 
\cite{anwar2019applications,annessi2017s,narula2018requirements}.
To this end, \paisa includes a {\it time synchronization} technique: at boot time, \iotdev  
synchronizes with a trusted server managed by the device manufacturer. See 
Sections~\ref{subsec:protocol-overview} and \ref{subsec:protocol} for details.

To summarize, \paisa is comprised of all aforementioned components.
Figure \ref{fig:iotdev-flow} presents a high-level overview of \paisa workflow. 
As soon as \iotdev boots, it synchronizes its time with the manufacturer server. 
Next, it attests its software and composes an announcement packet including the current timestamp, 
the attestation result, the device description, and a signature.
Then, \iotdev broadcasts the packet via WiFi. This is repeated for every timer interrupt, which is 
scheduled (likely configured by the manufacturer\footnote{It is debatable whether any other party
should be allowed to set the announcement schedule.}) according to the desired use-case. 
Each announcement is received by the \paisa app on every user device within range. After validating
the announcement, the app alerts the user to \iotdev's presence.

\section{System \& Adversary Models}\label{sec:system_assumption}
\subsection{Entities Involved}\label{subsec:entities}
\paisa considers three entities: \iotdev, \usrdev, and the manufacturer server (\mfr), 
which is responsible for provisioning \iotdev at production time.

\noindent{$\boldsymbol{\iotdev}$} is a resource-constrained IoT device installed either (1) in a public space, e.g., airports, restaurants
concert/sports venues, or stores, or (2) in a semi-private space, e.g., hotel rooms or Airbnb rentals.
\iotdev is assumed to be equipped with a TEE to protect \paisa TCB from untrusted software (including the OS).

\noindent {$\boldsymbol{\usrdev}$} is the personal and trusted device of the user. It is assumed to be within network transmission
range of \iotdev. \usrdev has an app that receives and verifies \paisa announcements.

\noindent {$\boldsymbol{\mfr}$} is a back-end (and sufficiently powerful) trusted server hosted by \iotdev manufacturer. 

\paisa assumes multiple \iotdev-s and multiple \usrdev-s in the same IoT-instrumented space, i.e., within network transmission 
range. \usrdev receives announcements from multiple \iotdev-s. \ \iotdev-s are unaware of \usrdev-s in their vicinity.
\paisa uses public key signatures to authenticate and verify announcements.
We assume a public-private key-pair (\iotdevpk, \iotdevsk) for each \iotdev and another key-pair (\mfrpk, \mfrsk) for each \mfr.
\mfrpk is used to authenticate \iotdev as part of announcement verification.

\begin{figure}[t]
  \centering
  \captionsetup{justification=centering}
  \includegraphics[width=1\columnwidth]{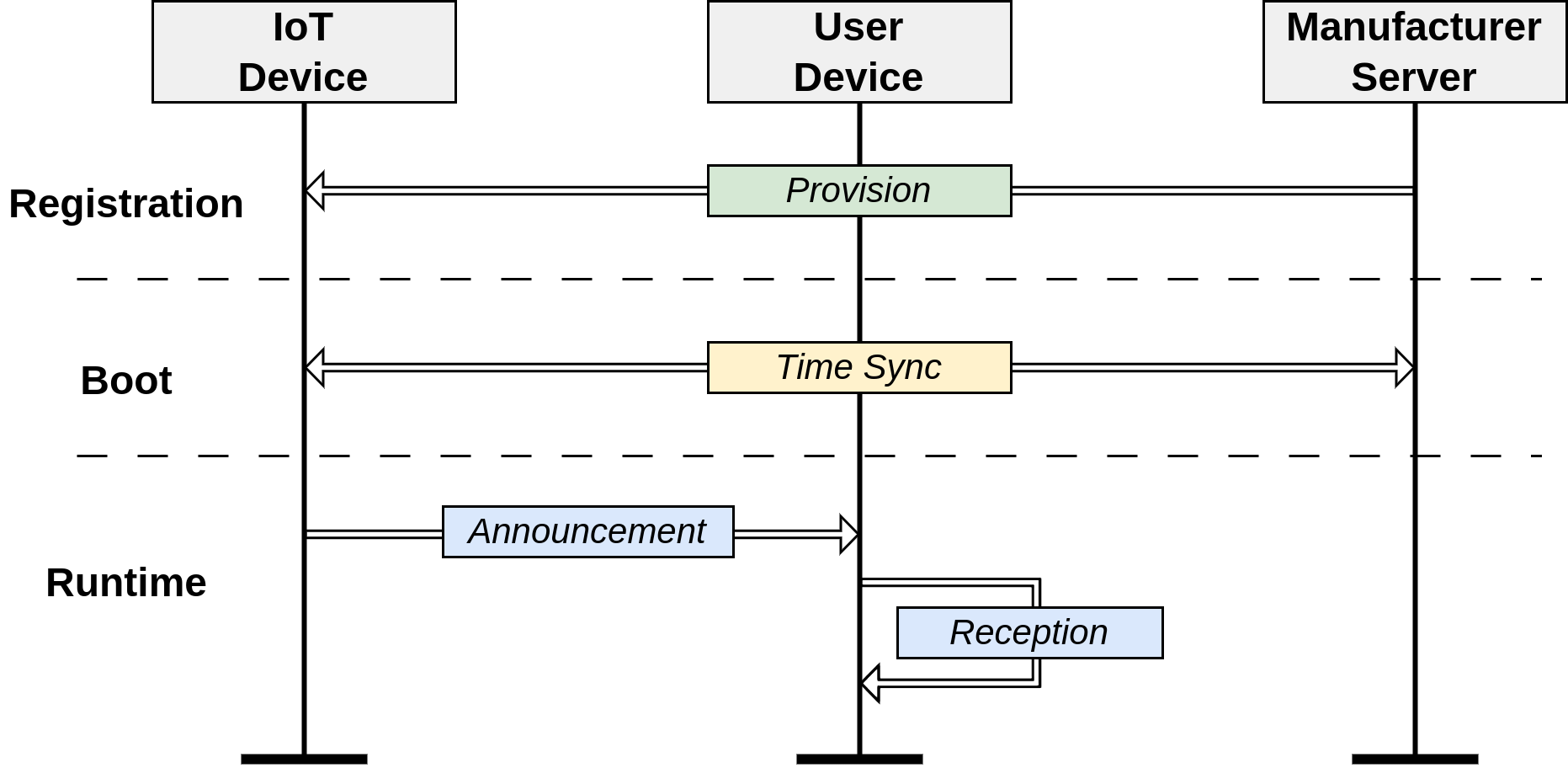}
  \vspace{-.3cm}
  \caption{\small \paisa Protocol Overview.}
  \label{fig:protocol-overview}
  \vspace{-.4cm}
\end{figure}

\subsection{\paisa Protocol Overview} \label{subsec:protocol-overview}
\paisa protocol has three phases: \registration, \boot, and \runtime. Figure \ref{fig:protocol-overview} shows its overview.

\noindent {$\boldsymbol{\registration}$} phase takes place when \iotdev is manufactured and provisioned.
At the time of the registration, besides installing software, \mfr installs \paisa TCB on \iotdev and 
provisions it with a device ID, a description, and a keypair (\iotdevpk, \iotdevsk) using \provision request.
Further details about the device description are in Section \ref{subsec:protocol}.
A provisioned \iotdev is eventually sold and deployed by its owner/operator.

\noindent {$\boldsymbol{\boot}$} phase is executed at \iotdev boot, after a reset or a power-on.
Before going into normal operation, \iotdev synchronizes its time with \mfr using \timesync 3-way protocol.
At the end of this phase, the initial announcement is generated.

\noindent {$\boldsymbol{\runtime}$} phase corresponds to \iotdev's normal operation. 
In this phase, \iotdev announces its presence based on a preset timer interval. 
Announcement periodicity is set by \mfr. (We are not advocating allowing owners to set this.) 
Whenever triggered by the timer, \announcement procedure is invoked. It attests \iotdev software and broadcasts an 
announcement (\announcemessage). A nearby \usrdev receives \announcemessage using its \reception app, which parses
and verifies \announcemessage. If the verification succeeds, \announcemessage is displayed to the user. 

For the complete protocol description, see Section \ref{subsec:protocol}.

\subsection{Adversary Model}\label{subsec:adversary}
We consider an adversary \sadv that has full control over \iotdev memory, including flash and RAM, except 
for the TCB and its data inside the TEE. \sadv can attempt to tamper with any \iotdev components and peripherals, 
including sensors, actuators, network interfaces, and debug ports, unless they are configured as secure by the TEE. 
All messages exchanged among \iotdev, \usrdev, and \mfr are subject 
to eavesdropping and manipulation by \sadv, following the well-known Dolev-Yao model \cite{DolevYao}. 
Furthermore, \registration phase is considered secure -- \mfr is trusted to correctly provision \iotdev 
and keep the latter's secrets. Also, \reception app on \usrdev is also considered trusted. \\

\noindent {\bf DoS Attacks:}
\sadv can essentially incapacitate ("brick") \iotdev by consuming all of its resources by malware. 
It can also keep all peripherals busy in an attempt to prevent \paisa TCB from broadcasting \announcemessage packets. 
It can ignore or drop outgoing packets or flood \iotdev with incoming malicious packets.
We also consider DoS attacks whereby a malware-controlled \iotdev reboots continuously and floods \mfr with frivolous 
\timesync requests. However, we do not consider \sadv that uses signal jammers to block \usrdev from receiving \announcemessage.
Such attacks are out of scope and there are techniques \cite{surveyjamming,jamming-wireless,jammer-sensornet} to prevent them.

\noindent {\bf Replay Attacks:}
we consider replay attacks whereby \sadv replays old/stale \announcemessage-s from any \paisa-compliant \iotdev-s.  
We also consider DoS attacks on \usrdev, e.g., \sadv replays 
old \announcemessage-s to swamp \usrdev network interface. 

\noindent {\bf Wormhole attacks:}\footnote{Replay and wormhole \announcemessage-s attacks overlap,
e.g., a replayed \announcemessage from a non-local \iotdev is both a replay and a wormhole attack.}
\paisa does not consider so-called wormhole attacks\cite{wormhole1,wormhole4}, whereby \sadv records 
and tunnels \announcemessage from remote locations (from outside \usrdev communication range).
There are well-known techniques \cite{wormhole1,wormhole2,wormhole3,surveydistancebounding} to tackle such attacks. 
However, \paisa provides to \usrdev coarse-grained location information, i.e., where \iotdev was manufactured and 
where it was deployed at \registration phase.

\noindent {\bf Physical Attacks:}
\paisa does not protect against physically invasive attacks on \iotdev, e.g., via hardware faults, modifying code 
in ROM, and extracting secrets via side-channels. We refer to \cite{ravi2004tamper} for protection against such attacks.
However, \paisa protects against non-invasive physical attacks, i.e., if \sadv tries to physically 
reprogram the device using wired debug interfaces such as JTAG. Such attacks are prevented using the secure boot feature 
of the TEE on \iotdev. 

\noindent {\bf Non-Compliant Devices:}
We do not consider attacks where \sadv physically infiltrates and deploys malicious (non-compliant) hidden devices in an 
IoT-instrumented space. As mentioned earlier, there are "spyware-type" techniques, such as \cite{mmwave,NLJD,bugdetector},
and other prior work, such as \cite{lumos,snoopdog}, that scan the area for hidden devices.
Albeit, even these techniques are error-prone, potentially computationally expensive, and time-consuming for users,
and/or require additional equipment.

\noindent 
{\bf Runtime Attacks:} Another limitation of \paisa is that it does not handle runtime control-flow attacks, 
such as buffer overflows, as well as non-control-flow and data-only attacks. \paisa can only detect software 
modifications via attestation. For mitigating these runtime attacks, there are techniques such as Control Flow Attestation 
(CFA) and Control Flow Integrity (CFI) \cite{litehax,cflat,tinycfa,oat,embedded-CFI,embedded-CFI-survey}.
Dealing with these attacks and deploying countermeasures is a good idea, though it is out-of-scope of this paper. 
Furthermore, many CFA/CFI techniques are resource-intensive, making their use challenging in IoT settings.

\subsection{Security \& Performance Requirements}\label{subsec:security_requirements}
Recall that the main objective of \paisa is to make \iotdev\ \privacyaware\, i.e., by guaranteed periodic announcements 
from \iotdev about its activity to adjacent  \usrdev-s, in the presence of \sadv defined in 
Section \ref{subsec:adversary}. To that end, \paisa must adhere to the following properties: 
\vspace{-0.1cm}
\begin{itemize}
    \item \emph{Unforgeability}: Announcements must be authenticated. \usrdev should be able to verify whether 
    \announcemessage is from a legitimate \iotdev, i.e., \sadv should not be able to forge \announcemessage.
    \item \emph{Timeliness}: Announcements must be released at fixed time intervals. \sadv should not be able to 
    prevent \announcemessage-s from being sent out.
    \item \emph{Freshness}: Announcements must be fresh and must reflect the current (software) health of 
    \iotdev. \sadv should not be able to launch replay attacks.
\end{itemize}  
\vspace{-0.1cm}
With respect to performance, \paisa must achieve the following:
\vspace{-0.1cm}
\begin{itemize}
    \item \emph{Low latency of \announcement}: 
    Announcements must be quick with minimal impact on \iotdev normal utility. 
    \item \emph{Low bandwidth of \announcement}: Announcements must be short to consume minimal 
    network bandwidth on \iotdev and \usrdev.  
\end{itemize}

\section{\paisa Design}\label{sec:design}
This section elaborates on the design and protocol overview presented in Sections \ref{sec:overview} and \ref{sec:system_assumption}.

\subsection{Design Challenges}\label{subsec:design_challenges}
There are a few design challenges (besides those mentioned in Section \ref{sec:overview}) to be addressed 
in order to achieve the security and performance requirements of \paisa.

\noindent \textbf{DoS Attacks Prevention on {$\boldsymbol{\iotdev}$}}: 
\sadv can launch DoS attacks by either keeping the MCU or the network peripherals busy, as mentioned in Section \ref{subsec:adversary}.
To prevent such attacks, \paisa configures both the timer and the network peripheral as {\em secure peripherals} controlled by the TEE.
By doing so, \paisa ensures that the MCU jumps into the TCB whenever the secure timer raises an interrupt according to scheduled periodicity.
Moreover, the timer interrupt is marked with the highest priority so that no other interrupt can preempt it.
This configuration (that determines which timer and network peripheral are trusted, and their interrupt priorities) 
is securely stored within the TEE. Hence, \sadv cannot tamper with it.
This also prevents DoS attacks that attempt to keep \iotdev from executing \paisa TCB that provides
guaranteed periodic broadcast of \announcemessage-s.

A typical target \iotdev might have 2-6 timers and multiple network peripherals, such as UART, SPI, and I2C on an MCU.
\paisa reserves one timer and one network peripheral for TCB usage.  
This means that the network interface (e.g., WiFi or BlueTooth) connected to that reserved network peripheral is marked as exclusive.
We admit that reserving a network interface exclusively for TCB use might be expensive for \iotdev, since at least one other interface 
(for regular use) would be needed. 

To address this issue, we implement a secure stub, akin to the ideas from \cite{rttee,minimuniotee,SeCloak}, to share the reserved network interface between secure and non-secure applications, detailed in Section \ref{subsec:impl_iotdev}.
For further discussion on this issue, see Section \ref{sec:limitation}.

\noindent \textbf{Bandwidth of {$\boldsymbol{\announcemessage}$}}: 
Broadcast messages are subject to size constraints that impact network efficiency and transmission capacity, regardless of the 
network type. Since the device description can be of arbitrary size, to minimize the size of \announcemessage, 
\paisa uses a fixed size broadcast message by placing all pertinent \iotdev information in a manifest file (\devmanifest).
\iotdev-generated \announcemessage-s carry only: (1) a URL that points to \devmanifest, and (2) some metadata: a timestamp, 
and a signature of \announcemessage. For the sake of simplicity, we assume that \devmanifest is hosted on \mfr.
\usrdev receives \announcemessage, verifies it, extracts the URL, and fetches \devmanifest from \mfr. 
Note that \devmanifest can also be hosted by other third parties or on a blockchain; its authenticity is based
on \mfr's signature at the time of provisioning.

\subsection{\paisa Protocol}\label{subsec:protocol}
Recall that \paisa includes three phases: \registration, \boot, and \runtime. 
Below we describe each phase in detail.

\begin{figure}[t]
  \centering
  \captionsetup{justification=centering}
  \includegraphics[width=1\columnwidth]{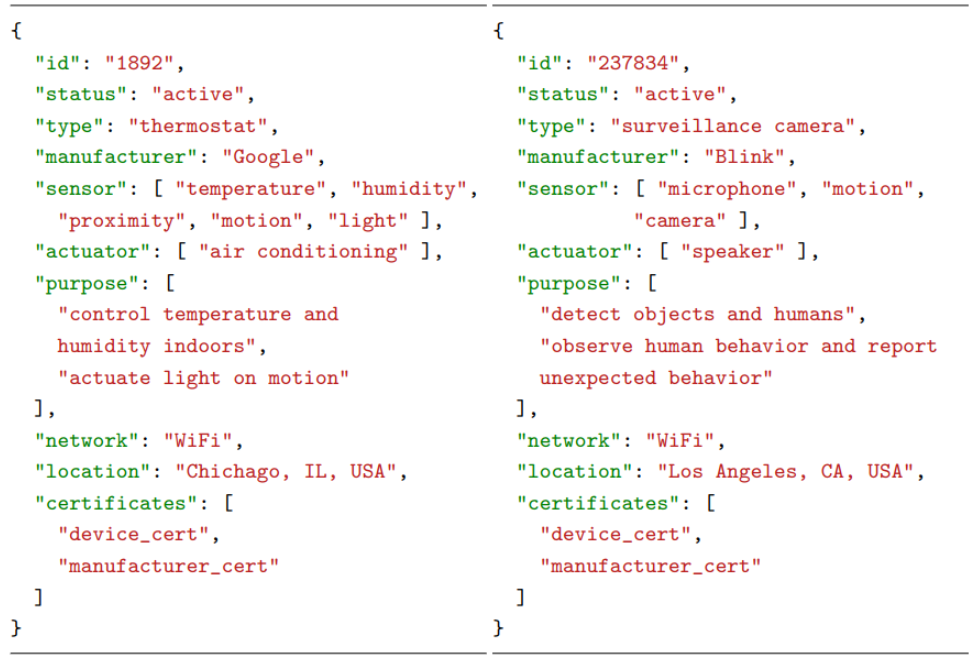}
  \vspace{-.3cm}
  \caption{\small Examples of \devmanifest. Left one is for Google Thermostat \cite{google-nest-thermostat} and right one is for Blink Security Camera \cite{blink-security-camera}.}
  \label{fig:dev_prf}
  \vspace{-.3cm}
\end{figure}

\begin{figure}
  \captionsetup{justification=centering}
  \begin{tcolorbox} [standard jigsaw, opacityback=0.3]
      \begin{protocol}\label{protocol:protocol-provisioning}
      \small
      \paisa Protocol at \registration consists of one procedure, \provision, and is realized as follows: \\
          
          \begin{compactenum} []
              \item {\textcolor{myblue}{\provision} $[\mfr \longrightarrow \iotdev]$:} \\
              Let (\mfrpk, \mfrsk) be \mfr keypair. \mfr provisions a \paisa-enabled \iotdev locally as follows: 
              \begin{compactenum}
                  \item \mfr compiles a \devmanifest including \iotdevid, the device description, and other metadata as shown in Figure \ref{fig:dev_prf}. \iotdevid \ is the identification number of \iotdev.
                  Also, \mfr hosts \devmanifest at \devmanifesturl location.
                  \item \mfr computes $\iotdevsofthash := \hashfunc(\iotdevsoft)$ and installs \iotdevsoft \ in the normal region of \iotdev, where \iotdevsoft \ is device software that is running during normal device operation and \hashfunc \ is a hash function.
                  \item \mfr generates $T$ as per equation \ref{eq:T-prov} and installs T in the secure region of \iotdev contained within the TEE.
                  \begin{equation}\label{eq:T-prov}
                     T := (\paisatcb||\iotdevid||\iotdevsofthash||\mfrpk||\devmanifesturl||\devmanifesturlfull||\tsvalue{cur})
                  \end{equation}
                  , where \paisatcb \ is \paisa TCB software,  \tsvalue{cur} is the current timestamp, \devmanifesturlfull is the full URL of \devmanifesturl if the URL is shortened, and \devmanifesturl is the shortened URL.
                  \item \paisatcb \ in \iotdev picks a new keypair (\iotdevpk,\iotdevsk), stores \iotdevsk, and outputs \iotdevpk to \mfr. 
                  \item \mfr computes $\devmanifestsig:=\sigfunc(\mfrsk,\devmanifest)$, where \sigfunc \ is a signature function, and appends \devmanifestsig and \mfrpk to \devmanifest hosted at \devmanifesturl.
              \end{compactenum}
          \end{compactenum}
      \end{protocol}
  \end{tcolorbox}
  \vspace{-.4cm}
  \caption{\small \registration Phase of \paisa.}
  \label{fig:prot-reg}
  \vspace{-.4cm}
\end{figure}

\subsubsection{Registration}\label{subsubsec:registration}
In this phase, \mfr interacts with \iotdev to provision it with secrets and information needed to
enable \paisa. Figure \ref{fig:prot-reg} depicts this phase.

\noindent {\bf Device Manifest:} \mfr creates \devmanifest for \iotdev, including device ID (\iotdevid), 
a description which includes:\footnote{This is just a sample list; some attributes 
might be optional and others might be needed.}
\begin{quote}
device type/model, manufacturer, date/location of manufacture, types of sensors/actuators, deployment purpose, 
network interfaces, owner ID, and location of deployment
\end{quote}
Figure \ref{fig:dev_prf} shows \devmanifest examples.
\devmanifest can also contain a link to \iotdev developer documentation, as mentioned in \cite{nist-reco}.
Note that, whenever the owner changes \iotdev's location, the corresponding manifest must be updated accordingly. 
The granularity of this location information influences the ability to mitigate wormhole attacks.
We believe that the contents of \devmanifest suffice to make a user aware of \iotdev capabilities.
However, the exact contents of \devmanifest are left up to the manufacturer. 

\mfr stores each \devmanifest it in its database and generates a publicly accessible link \devmanifesturl.
Since \devmanifesturl can be long, we recommend using a URL shortening service (such as \verb+Bitly+\cite{bitly} or 
\verb+TinyURL+\cite{tinyurl}) to keep \devmanifesturl short and of fixed size. 

Hereafter, we use \devmanifesturl to denote the short URL and \devmanifesturlfull -- the original URL. 
(Note that if the shortening service is not used, then \devmanifesturl is identical to \devmanifesturlfull.)

For simplicity's sake, besides manufacturing \iotdev, we assume that \mfr is responsible for deploying and maintaining the 
software (\iotdevsoft)  on \iotdev. However, in practical scenarios, other entities, such as software vendors, can be 
involved in managing individual applications on \iotdev.  In such cases, vendors must be integrated into the trust-chain 
by including their information and certificates into \devmanifest.
Whenever a vendor-imposed software update occurs, \devmanifest must be updated and re-signed by \mfr.
We further discuss this update process in Section \ref{sec:limitation}.

\noindent {\provision:} \mfr installs \iotdevsoft\ and \paisa TCB (\paisatcb) into the normal and 
secure regions of \iotdev, respectively. \mfr ensures that the timer and the network peripheral 
are configured as secure and exclusively accessible to \paisatcb.
Also, \mfr sends \iotdevid\ and a hash of \iotdevsoft\ to \iotdev to be stored in \paisatcb.
Next, \paisatcb\ picks a new public/private key-pair (\iotdevpk, \iotdevsk) and sends \iotdevpk to \mfr 
for certification. 
\mfr also gives the current timestamp to \paisatcb, to be used for implementing a clock on \iotdev (see Section \ref{subsubsec:boot}).
\mfr appends \iotdevpk\ and the hash of \iotdevsoft\ to \devmanifest. 

Finally, to authenticate \devmanifest, \mfr signs \devmanifest using \mfrsk\ and appends the 
signature and its own certificate to \devmanifest. 
Alternatively, \mfr could directly register \iotdevpk\ with a Certificate Authority (CA) 
if there is a suitable deployed public key infrastructure (PKI), and include \iotdev's certificate in \devmanifest.

Also, \devmanifesturlfull is included in \devmanifest\ so that \usrdev, when it 
later uses \devmanifesturl, can detect if the redirection is wrong. Also, for sanity purposes, 
\mfr can include a "status" flag in \devmanifest to indicate if \iotdev is revoked, e.g., reported stolen.

\begin{figure}
  \captionsetup{justification=centering}
  \begin{tcolorbox} [standard jigsaw, opacityback=0.3]
      \begin{protocol}\label{protocol:protocol-boot}
          \small
          \paisa\ \boot consists of one procedure, \timesync, and is realized as follows. \\      
          \begin{compactenum}[]
            \item { \textcolor{myblue}{\timesync} $[\iotdev \longleftrightarrow \mfr]$:}
                Assume a map $Map_{\svr} := <\iotdevid, \tsvalue{\iotdev}>$ maintained by \mfr, where \iotdevid \ is  
                ID of \iotdev provisioned using \provision and \tsvalue{\iotdev} is the latest registered timestamp of 
                \iotdev.  \timesync is defined by three interactions [\syncreq, \syncresp, \syncack]:
                  \begin{compactenum}
                    \item \textcolor{burgundy}{\syncreq} $[\iotdev \longrightarrow \mfr]$ : \\
                        When \iotdev boots:
                      \begin{compactenum}
                        \item Computes $\syncreq := (\iotdevid,  \noncevalue{\dev}{1}, \tsvalue{prev}, \sigvalue{Req})$, 
                        where \noncevalue{\dev}{1} is a nonce, \tsvalue{prev} is the previous timestamp, and 
                        \begin{equation}\label{eq:sig-req}
                            \sigvalue{Req} := \sigfunc(\iotdevsk,\hashfunc(\iotdevid||\noncevalue{\dev}{1}||\tsvalue{prev}+1))
                        \end{equation}
                        \item Sends $\syncreq$ to \mfr.
                      \end{compactenum}
                    
                    \item \textcolor{burgundy}{\syncresp}  $[\iotdev \longleftarrow \mfr]$ : \\
                    Upon receiving \syncreq, \mfr:
                      \begin{compactenum}
                        \item Checks if \ $\tsvalue{prev} + 1$ is consistent with the latest registered timestamp in 
                        $Map_{\svr}$. If fails, outputs $\perp$ and ignores \syncreq.
                        \item Verifies $\sigvalue{Req}$ using \iotdevpk. If fails, outputs $\perp$ and ignores \syncreq; 
                        otherwise, continues.
                        \item Computes $\syncresp := (\iotdevid, \noncevalue{\dev}{1}, \noncevalue{\svr}{1}, \tsvalue{cur}, 
                        \sigvalue{Resp})$, where \noncevalue{\svr}{1} is a nonce and \tsvalue{cur} is the 
                        current timestamp of \mfr, and 
                        \begin{equation}\label{eq:sig-resp}
                            \sigvalue{Req} := \sigfunc(\mfrsk,\hashfunc(\iotdevid||\noncevalue{\dev}{1}|| 
                            \noncevalue{\mfr}{1}|| \tsvalue{cur}))
                        \end{equation}
                        \item Sends $\syncresp$ to \iotdev.
                      \end{compactenum}
                    \item \textcolor{burgundy}{\syncack} $[\iotdev \longrightarrow \mfr]$ : \\
                    Upon receiving \syncresp, \iotdev:
                      \begin{compactenum}
                        \item Verifies $\syncresp$ using \mfrpk. If fails, outputs $\perp$, ignores \syncresp, and 
                        repeats \timesync; otherwise continues.
                        \item Sets $\tsvalue{prev} := \tsvalue{cur}$ from \syncresp.
                        \item Computes $\syncack := (\iotdevid, \noncevalue{\dev}{2}, \noncevalue{\svr}{1}, \tsvalue{prev}, 
                        \sigvalue{Ack})$, where \noncevalue{\dev}{2} is a nonce of \iotdev and
                        \begin{equation}\label{eq:sig-ack}
                            \sigvalue{Ack} := \sigfunc(\iotdevsk,\hashfunc(\iotdevid||\noncevalue{\dev}{2}|| 
                            \noncevalue{\mfr}{1}|| \tsvalue{prev}))
                        \end{equation}
                        \item Sends $\syncack$ to \mfr.
                      \end{compactenum}
                  \end{compactenum}
                  Finally, \mfr verifies $\sigvalue{Ack}$ with \iotdev. If successful, \mfr stores $\tsvalue{prev}$ as the latest registered timestamp of \iotdev.
          \end{compactenum}
      \end{protocol}
  \end{tcolorbox}
  \vspace{-.4cm}
  \caption{\small \boot Phase of \paisa}
  \label{fig:prot-boot}
  \vspace{-.4cm}
\end{figure}

\begin{figure*}
  \captionsetup{justification=centering}
  \begin{tcolorbox} [standard jigsaw, opacityback=0.3]
      \begin{protocol}\label{protocol:protocol-runtime}
          \small
          \paisa runtime consists of two procedures: \announcement and \reception: \\       
          \begin{compactenum}[]
            \item {\textcolor{myblue}{\announcement} $[\iotdev \longleftrightarrow \usrdev]$:} 
                  Let \tsvalue{\dev} be \iotdev clock realized using a secure timer and the latest timestamp received via 
                  \timesync. \announcement is defined by two sub-procedures [\attest, \announce]. Also, let \attestinterval and \announceinterval 
                  be the periodicity of \attest and \announce, respectively.  
                  \begin{compactenum}
                    \item \textcolor{burgundy}{\attest} $[\iotdev \longrightarrow \iotdev]$ : \\
                        If \ $\tsvalue{\dev}\ \%\ \attestinterval == 0$, \iotdev generates an attestation report:
                      \begin{compactenum}
                          \item Measures program memory: $Att_{\dev} := \hashfunc(\iotdevsoft)$.
                          \item Sets $\attestresult := 1$ if $Att_{\dev} == \iotdevsofthash$, where \iotdevsofthash is the expected 
                          hash of \iotdev software installed during \provision. Otherwise, $\attestresult = 0$.
                          \item Outputs $\attestreport := (\attestresult, \tsvalue{\dev})$, where \tsvalue{\dev} is the timestamp when 
                          the attestation report is generated.
                      \end{compactenum}
                    \item \textcolor{burgundy}{\announce} $[\iotdev \longrightarrow \usrdev]$ : \\
                      If \ $\tsvalue{\dev}\ \%\ \announceinterval == 0$, \iotdev broadcasts an announcement packet:
                      \begin{compactenum}
                          \item Generates $\announcemessage := 
                          (\noncevalue{\dev}{},\tsvalue{\dev},\devmanifesturl,\attest,\sigvalue{\anno})$, 
                          where \noncevalue{\dev}{} is a nonce, \tsvalue{\dev} is the current timestamp, \devmanifesturl is the stored 
                          link pointing to \devmanifest given at \provision, and 
                          \begin{equation}\label{eq:sig-anno}
                              \sigvalue{\anno} := \sigfunc(\iotdevsk, \hashfunc(\iotdevid||\noncevalue{\dev}{}||\tsvalue{\dev}||\devmanifesturl||\attest))
                          \end{equation}
                          \item Broadcasts \announcemessage.
                      \end{compactenum}
                \end{compactenum}
              \item {\textcolor{myblue}{\reception} $[\usrdev \longleftrightarrow \usrdev]$:} 
                  \usrdev maintains a timer \tsvalue{\usrdev} synchronized with the world clock. Upon receiving \announcemessage from a \iotdev, \usrdev executes \reception.
                  \reception is defined by a sub-procedure [\verifymessage] :
                  \begin{compactenum}
                    \item Parses \announcemessage and extracts: $(\tsvalue{\dev},\devmanifesturl,\attest,\sigvalue{\anno})$. 
                    Next, fetches \devmanifest from \devmanifesturl.
                    \item \textcolor{burgundy}{\verifymessage} $[\usrdev \longrightarrow \usrdev]$ : \\
                     Upon receipt of \devmanifest, verifies \announcemessage:
                      \begin{compactenum}
                          \item Checks if $(\tsvalue{\usrdev} - \epsilon) < \tsvalue{\dev}$, where $\epsilon$ is the tolerance 
                          delay window. If not, discards and outputs $\perp$.
                          \item Retrieves \devmanifestsig and \mfrpk from \devmanifest, and verifies \devmanifestsig using \mfrpk. 
                          If fails, aborts and outputs $\perp$.
                          \item Retrieves \iotdevpk and verifies \sigvalue{\anno}. If fails, aborts and outputs $\perp$.
                      \end{compactenum}
                    \item Outputs (\devmanifest, \attest).
                \end{compactenum}
            \end{compactenum}
      \end{protocol}
  \end{tcolorbox}
  \vspace{-.4cm}
  \caption{\small \runtime Phase of \paisa}
    \label{fig:prot-runtime}
    \vspace{-.4cm}
\end{figure*}

\subsubsection{\boot}\label{subsubsec:boot}
As mentioned earlier, \announcemessage must contain the timestamp of \mfr to prevent replay attacks.
Some IoT devices feature a reliable real-time clock (RTC)\cite{rtc-example} 
powered by a separate power source, thus ensuring that \iotdev time is always accurate.
However, most resource-constrained IoT devices lack such an RTC.
To this end, \paisa includes a secure time synchronization (\timesync) protocol between \iotdev and \mfr.
It assumes that \mfr is both reachable and available at all times.

The main idea of \timesync is to receive the latest timestamp from \mfr whenever \iotdev (re)boots, or 
(optionally) at regular intervals. Figure \ref{fig:prot-boot} shows the \boot protocol. 

\noindent {\timesync:} After completing the boot-up sequence, \iotdev sends a time synchronization request 
\syncreq to \mfr, which includes \iotdevid \ and the previous timestamp \tsvalue{prev} given by \mfr at 
\provision or \timesync of the last boot. \syncreq also contains a signature to authenticate its origin
as a legitimate \iotdev, and prevent DoS attacks on \mfr via flooding of fake requests.\footnote{We acknowledge
that signature itself might be a DoS attack vector since it consumes \mfr's resources to verify.}
Upon receiving \syncreq, \mfr verifies the signature using \iotdevpk and responds with  \syncresp 
that includes the current timestamp \tsvalue{cur}.
Upon receipt of a \syncresp, \iotdev verifies the signature using \mfrpk obtained at \provision.
If verification succeeds, \iotdev updates its local timestamp and sends an authenticated acknowledgment 
\syncack to \mfr. Finally, \mfr verifies \syncack and updates its local registered time database for 
\iotdevid. Next time \iotdev requests a \timesync, \mfr will know whether the signature is based on 
the same \tsvalue{prev} it previously sent. At the end of the protocol, \iotdev and \mfr have the same 
\tsvalue{cur}. Given the unavoidable network transmission latency, 
we suggest keeping a window of acceptance $\epsilon$ when verifying timestamps.

Subsequently, \iotdev can be synchronized with \mfr by re-starting the 
secure timer after receiving and updating \tsvalue{prev}.
Thereafter, \iotdev computes the latest time by adding \tsvalue{prev} and the 
secure timer value; we denote this time as \tsvalue{\dev}.
However, since this secure timer value might still deviate due to hardware inconsistencies, 
repeating \timesync at regular intervals is recommended.

\subsubsection{Runtime}\label{subsubsec:runtime}

The current \paisa design uses a {\em push} model, whereby 
\iotdev periodically transmits \announcemessage-s at fixed intervals. 
An intuitive alternative is to use a {\em pull} \ model, in which \usrdev announces its
presence first and, in response, solicits information from all nearby \iotdev-s. 
This is similar to the Access Point (AP) discovery process in WiFi:  \usrdev 
emits a ``Probe Request'' to which an AP responds with a ``Probe Response'' 
containing information about the various network parameters to establish the connection.
In the same fashion, \iotdev that receives a ``Probe Request'' 
could include \announcemessage in the ``Probe Response'' and send it to \usrdev.
One advantage of the pull model is that \announcemessage-s are only sent when
they are needed, thus reducing the burden on individual \iotdev-s and easing the network traffic congestion. 
On the other hand, it becomes more challenging to deal with ``sleeping'' or intermittently powered-off \iotdev-s, thereby raising the energy consumption issues. 
In any case, we intend to explore the pull model further as part of near-future work.

\paisa runtime shown in Figure \ref{fig:prot-runtime} involves two procedures: (1)  
\announcement on \iotdev is part of \paisatcb, installed at \provision time, and (2)  
\reception is an app on \usrdev, installed by the user.

\noindent {\announcement:} \paisa implements two time intervals using secure timer on \iotdev, \attestinterval and
\announceinterval, which govern when \attest and \announce must be executed, respectively, triggered by the timer interrupt.
During \attest, i.e., when \tsvalue{\dev} matches \attestinterval, \paisa measures \iotdev memory containing 
\iotdevsoft \ and compares it with the hash of \iotdevsoft \ stored at \provision time.
If the measurements match, \iotdev sets $\attestresult=true$ and
\attestreport =  ($\attestresult, \tsvalue{\dev})$ and stores the latter in secure RAM.

During \announce, i.e., when \tsvalue{\dev} matches \announceinterval, \iotdev generates 
new \announcemessage composed of: a nonce, the current timestamp \tsvalue{\dev}, 
\devmanifesturl given at \provision time, \attestreport from the latest attestation as per \attestinterval, and a signature over its content.
The size of \announcemessage depends on the signature algorithm used. 
Also, whenever the \devmanifest or \devmanifesturl is updated (e.g., software update, 
maintenance shutdown, or a change of the shortened URL,), \mfr sends the updated 
\devmanifesturl to \iotdev at the time of \timesync.

\noindent {\bf \attest and \announce periodicity:}
If \attestinterval is the same as \announceinterval, then attestation and announcement are performed sequentially.
This is recommended so that \usrdev always receives the latest information about \iotdev.
However, periodicity can be adjusted based on device capabilities and desired use-cases. 
If \iotdev is a weak low-end device and/or must prioritize its normal applications, \attestinterval can 
be longer than \announceinterval.\footnote{For example, \attestinterval can be set to one day while \announceinterval -- 
to 10 seconds, implying that \usrdev can confirm that \iotdev exists in the locality and that it is not 
compromised, at least, the last 24 hours, provided that the verification is successful.}
In our experiments, \attest time is much smaller than \announce time because signing 
takes more time than just hashing a small amount of memory. 

\noindent {\reception:} After receiving \announcemessage from \iotdev, \usrdev first parses it and checks 
if the received \tsvalue{\dev} is within $[\tsvalue{\usrdev} - \epsilon, \tsvalue{\usrdev}]$, where 
\tsvalue{\usrdev} is the clock value of \usrdev, and $\epsilon$ is the toleration delay window of the assumed 
network. If \announcemessage is fresh, then \usrdev fetches \devmanifest from the link \devmanifesturl and 
verifies \devmanifest based on the public key \mfrpk and the signature \devmanifestsig embedded in \devmanifest.
Next, it verifies the signature of \announcemessage with the public key of \iotdev, also embedded in \devmanifest.
Upon successful verification of the signatures, \usrdev acknowledges the legitimacy of the announcement source, 
thereby confirming that the corresponding \iotdev is in its network reach.
Furthermore, by reading \attest, \usrdev learns whether \iotdev is a trustworthy state since the last attestation.
If \attest fails, \usrdev disregards \announcemessage and alerts the user of a potentially compromised \iotdev.

We note that user linkage might occur if \usrdev fetches multiple \devmanifest-s from the same \mfr, 
assuming the latter is honest-but-curious. To mitigate this, there are well-known techniques for 
anonymous retrieval, such as Tor. Although this issue is somewhat outside the scope of this paper, 
we discuss it further in Section \ref{sec:limitation}.

\section{Implementation}\label{sec:implementation}
This section describes \paisa implementation details. 
All source code is publicly available at \cite{paisa-code}.

\subsection{Implementation Setup}\label{subsec:impl_setup}
As \iotdev, we use NXP LPC55S69-EVK \cite{nxpboard} development board, based on ARM Cortex-M33 MCU (in turn
based on ARMv8-M architecture) equipped with ARM TrustZone-M (TZ-M). The board runs at 150 MHz with 
640KB flash and 320KB SRAM. For the network interface, we connect a ESP32-C3-DevKitC-02 \cite{espboard} 
board, via UART to the NXP board. This network interface runs 2.4 GHz WiFi (802.11 b/g/n) and it 
is connected to the internet via a local router.

\mfr is emulated using a Python application running on a Ubuntu 20.04 LTS desktop with an Intel i5-11400 processor 
at 2.6GHz with 16GB RAM. \mfr is connected to \iotdev using UDP for \timesync.

\begin{figure}[t]
  \centering
  \captionsetup{justification=centering}
  \includegraphics[width=1\columnwidth]{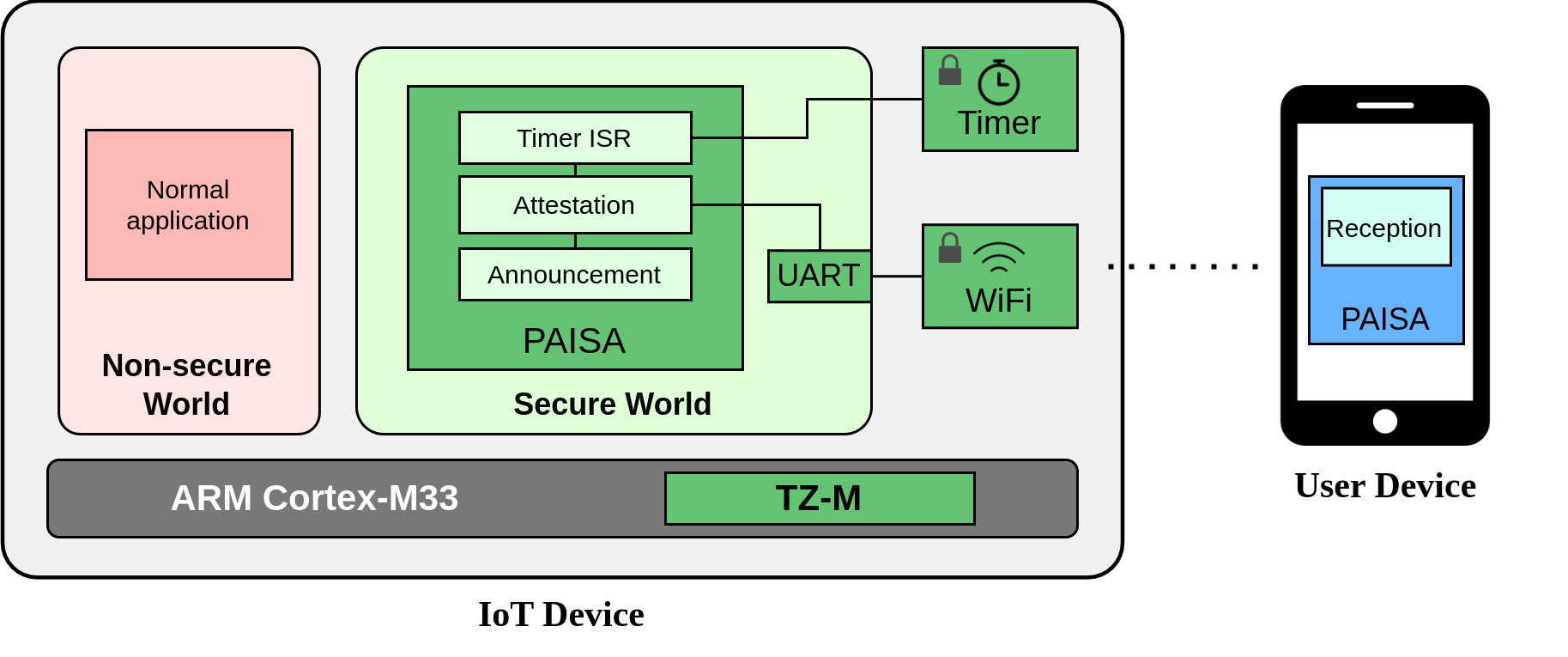}
  \vspace{-.3cm}
  \caption{\small \paisa Implementation Setup.}
  \label{fig:dev_config}
  \vspace{-.3cm}
\end{figure}

As \usrdev, we use a Google Pixel 6 \cite{pixel6}, with 8 cores running at up to 2.8GHz, which is used for \usrdev.
Both \iotdev and \usrdev use WiFi as their network interface to transmit/receive announcements.
Figure \ref{fig:dev_config} depicts the implementation architecture and Figure \ref{fig:paisa-poc} illustrates 
the complete prototype.

\noindent {\bf TCB configuration on TZ-M:}
CTIMER2 and UART4 peripherals are configured as secure, ensuring that only TCB can access them. 
This assurance is provided by TZ-M hardware, which raises a SecureFault (i.e., a hardware fault) whenever a non-secure 
application attempts to modify the configuration or access the secure peripherals directly.
When a SecureFault is issued, the MCU enters into the SecureFault handler within the TCB, where \paisa 
resets the MCU. Therefore, even if \sadv attempts to cause a DoS attack by raising SecureFaults, 
\paisa issues announcements by transmitting new \announcemessage as soon as the device awakes, before 
any normal activity. Also, the secure timer is configured, with the highest priority, to interrupt the MCU 
via the NVIC every \announceinterval. Hence, no other user-level interrupt can preempt the announcement schedule.

\subsection{Implementation Challenges}\label{subsec:impl_challenges}
\noindent \textbf{How to announce?} 
An interesting challenge is how to broadcast \announcemessage when \usrdev does not have a connection with \iotdev. 
A naive option is to broadcast \announcemessage via UDP packets.
However, this is not a robust model, since the local WiFi router in the subnet must be trusted to relay 
packets to \usrdev-s. Moreover, it requires \usrdev-s to be connected to the router to receive 
\announcemessage-s, which is not a fair assumption.
To mitigate this issue, we use the IEEE 802.11 standard WiFi Beacon Frames \cite{wifibeacon}. 
Beacon frames are typically used by routers or APs to advertise their presence. 
\paisa can implement such beacon frames to broadcast its \announcemessage letting other 
devices know \iotdev presence, akin to a router.
More specifically, \paisa uses vendor-specific elements in the beacon frame to populate \announcemessage.

\noindent \textbf{\announcemessage size limitation:} 
\announcemessage size is limited to 255 bytes as per the length of a vendor-specific element in a beacon frame. 
Hence, to fit into that size, we minimized all fields in \announcemessage. 
By using \verb+Bitly+, \devmanifesturl can be reduced to 11 bytes. 
By using ECDSA with Prime256v1 curve, \sigvalue{\anno} can be reduced to 64 bytes.
By using the UNIX Epoch format, \tsvalue{\dev} requires only 4 bytes. 
Only 5 bytes are needed for the attestation report, including one byte for the attestation result (a boolean) and 
4 bytes for the attestation timestamp.
In total, \announcemessage size is about 116 bytes including a 32-byte nonce. 

A typical WiFi router beacon frame observed in our experiments is between 200 and 450 bytes.
The beacon frame generated by \paisa\ \announcemessage is 240 bytes. 
It is relatively small since it contains only one vendor-specific element
and no other optional tags (besides required fields), in contrast with a typical beacon frame 
that carries multiple proprietary optional tags.

\noindent \textbf{Signing overhead:} Computing a signature is performance-intensive.
Some very low-end devices cannot even afford them due to heavy cryptographic computations, 
and some take several seconds to do so. Fortunately, TEEs such as TrustZone, are (although optional) 
usually  equipped with cryptographic hardware support.
In our implementation, we use the cryptographic accelerator, CASPER, on the NXP board to perform Elliptic 
Curve Cryptography (ECC) to reduce signing overhead.

\subsection{Trusted Software in \iotdev}\label{subsec:impl_iotdev}
Figure \ref{fig:dev_config} shows that \iotdev contains three applications: non-secure application 
in the normal region, \paisa TCB in the secure region, and network stack connected to the secure 
UART4 interface. 

\noindent {\bf Non-secure application:} We implemented a sample thermal 
sensor software as a non-secure application in the normal region.
The software reads temperature data from the sensor (on the NXP board) every second and sends it to an external 
server via the network interface. Since the network interface is exclusive to the secure world, we implemented a 
secure stub that can be invoked by an NSC function, allowing non-secure applications to access the network interface. 
This stub always prioritizes \paisa announcements over other requests.

For cryptographic operations, we use Mbed TLS library \cite{mbedtls} on both \iotdev and \mfr.
At \provision, \iotdev and \mfr both sample new pairs of ECC keys based on the Prime256v1 curve.

\noindent {\bf $\boldsymbol{\paisa}$ TCB} mainly contains three modules: Secure Timer ISR, Attestation, and Announcement. 
Secure Timer ISR, connected to CTIMER2, is executed when the announcement interval \announceinterval is triggered 
via the NVIC. This ISR first calls Attestation module, if \attestinterval is met, and then invokes Announcement module.
Attestation module computes SHA256 over application program memory, in 4KB chunks, and generates 
\attestreport, as shown in Figure \ref{fig:prot-runtime}.
Next, Announcement module creates \announcemessage and sends it to the WiFi interface using \verb+USART_WriteBlocking()+.

\begin{figure}[t]
  \centering
  \captionsetup{justification=centering}
  \includegraphics[width=1\columnwidth]{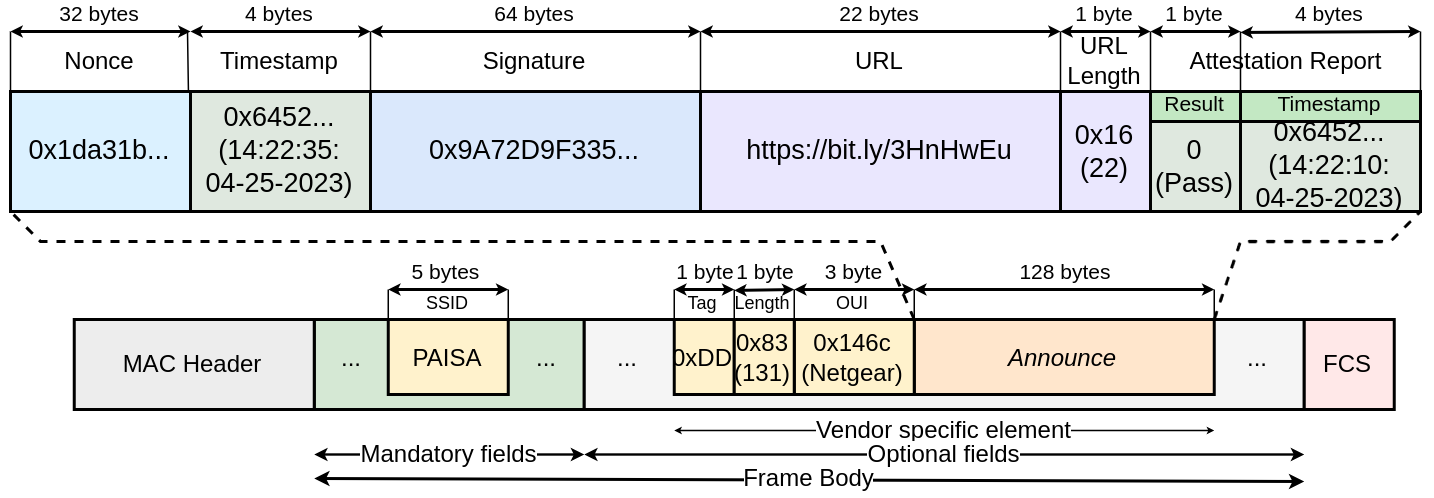}
  \vspace{-0.3cm}
  \caption{\small Example of \announcemessage.}
  \label{fig:announce-with-data}
  \vspace{-0.3cm}
\end{figure}

\noindent {\bf Network Stack:}
The ESP32-C3-DevKitC-02 board houses WiFi and Bluetooth on a single board, running on a 32-bit RISC-V single-core processor 
running at 160 MHz. The board complies with IEEE 802.11b/g/n protocol and supports Station mode, SoftAP mode, 
and SoftAP + Station mode. \paisa TCB uses Station mode for \timesync with \mfr and SoftAP mode for \announcement 
to \usrdev.

After receiving \announcemessage via \verb+uart_read_bytes()+, WiFi module generates a beacon frame 
using \verb+esp_wifi_80211_tx()+ API and sets \verb+SSID="PAISA"+. 
Figure \ref{fig:announce-with-data} shows an example beacon frame produced.
It includes \announcemessage in the vendor-specific element: first byte $(0xdd)$
indicates Element ID, second byte $(0x83)$ denotes length of the tag, and next three bytes 
$(0x00,0x14,0x6c)$ represent Organizationally Unique Identifier (OUI) for Netgear, while
remaining bytes carry \announcemessage contents. The beacon frame is transmitted according to the 
same WiFi beacon standard.

\subsection{\reception App in \usrdev}\label{subsec:impl_usrdev}
We implemented \reception as an Android app on \usrdev -- Google Pixel 6. It was  
developed using Android Studio. To scan for beacon frames, \reception requires location and 
WiFi access permissions enabled by setting \ \verb+ACCESS_FINE_LOCATION+ \ and \ \verb+CHANGE_WIFI_STATE+ 
in the app configuration.

\begin{figure}[h]
  \centering
  \captionsetup{justification=centering}
  \includegraphics[width=.9\columnwidth]{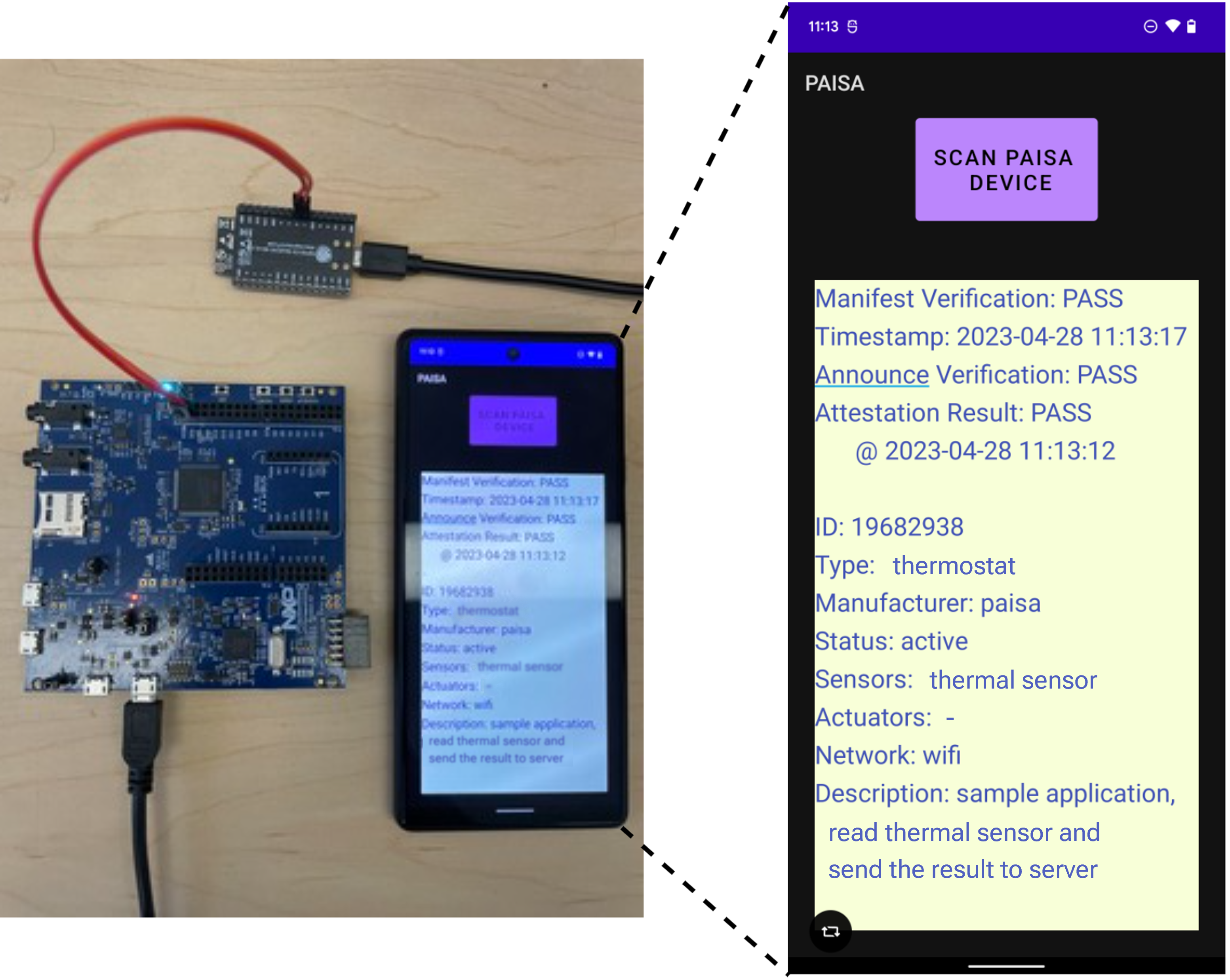}
  \vspace{-0.3cm}
  \caption{\small \paisa Proof-of-Concept. The Phone screenshot on the right side shows \reception app with device details of \iotdev \\ (emulated on the NXP board beside it).}
  \label{fig:paisa-poc}
  \vspace{-0.3cm}
\end{figure}

\reception uses \verb+getScanResult()+ API in \verb+wifi.ScanResult+ library to scan and identify WiFi 
beacon frames containing \verb+SSID=+ \verb+"PAISA"+. Then, it uses \verb+marshall()+ API from \verb+os.Parcel+ library 
to extract the list of vendor-specific elements from the frame.
Next, the app parses \announcemessage and fetches \devmanifest from \devmanifesturl using \verb+getInputStream+ 
API in \verb+net.HttpURLConnection+ library.
After receiving \devmanifest, it verifies signatures in \devmanifest and \announcemessage using the corresponding 
public keys via \verb+java+. \verb+security+ library.
Finally, it displays the device description and the attestation report on \usrdev screen, as shown in 
Figure \ref{fig:paisa-poc}. \reception app also has \verb+"SCAN PAISA DEVICE"+ button (as shown in the figure) 
to explicitly scan for \iotdev.

\section{Evaluation}\label{sec:evaluation}
This section presents the security and performance analysis of \paisa. 

\subsection{Security Analysis}\label{sec:security_analysis}

We argue the security of \iotdev by showing an \sadv (defined in Section \ref{subsec:adversary}) that attempts to attack \timesync and \announcement modules, and how \paisa defends against such \sadv.

\sadv who controls the normal region of \iotdev, can attack \paisa in the following ways: (a) attempt to modify the code, data, and configuration of the secure modules, or try to read \iotdevsk (b) attempt to keep normal application busy (for e.g., by running an infinite loop), (c) attempt to continuously raise interrupts to escalate into the privileged mode of execution in the normal region, (d) attempt to broadcast fake or replay old \announcemessage-s, (e) tamper with or drop \timesync messages, and (f) attempt to leak privacy of \usrdev.

First, the TZSC in TZ-M hardware ensures the protection of all memory within the secure region including the secure peripheral configuration.
Thus, it raises a SecureFault when (a) occurs and gives control back to the secure region handler. 

Second, the NVIC configuration of MCU ensures that the secure timer has the highest priority (i.e., not preemptible), and when that timer interrupt occurs, it guarantees to invoke the secure timer ISR within the secure region. 
Hence, despite \sadv attempts to block announcements by (b) or (c), \announcement is executed in a timely manner.
Moreover, the network module is under the control of secure UART, thus, even that cannot be blocked by malicious applications.
Additionally, since the announcements reach \usrdev within one hop, \sadv on the internet is totally harmless.

Third, the unforgeability guarantee of signature schemes ensures that \sadv cannot generate a correct \announcemessage without knowing \iotdevsk. 
This entails, \sadv cannot modify the \attest report to hide compromised applications, modify the timestamp of old \announcemessage to create fake new ones, or make a \announcemessage point to a wrong \devmanifest; as \usrdev catches these during \verifymessage.
And similarly, \sadv cannot get away with replaying old \announcemessage with valid \attest report because \usrdev detects obsolete messages based on the timestamp in it.
Hence, (d) is not possible.

Fourth, messages exchanged in \timesync are all authenticated with signatures, so tampering is not viable. 
Next, since the network module on \iotdev is secure, \sadv cannot drop packets going out of \iotdev. 
However, \sadv on the internet can intercept and drop messages that are in transit between \iotdev and \mfr. 
For that, \paisa carefully retransmits when necessary as mentioned in Section \ref{subsec:protocol}. 
Additionally, \sadv can launch network DoS attacks by flooding \mfr or \iotdev during \timesync. 
Nonetheless, this does not harm the purpose of \paisa because, in that case, \iotdev did not even boot to resume its activity, so no need to announce \announcemessage anyway.

Lastly, \sadv compromising one or more \iotdev can attempt to trace \usrdev location. 
However, by virtue of PKC, \usrdev need not connect to any \iotdev to learn about the IoT activity in the vicinity. 
Therefore, there is no user privacy leakage at all.

The above five points conclude the security argument of \paisa, ensuring it meets all security requirements stated in Section \ref{subsec:security_requirements}.

\subsection{Performance Analysis}
Note that we measure the mean and standard deviation of each performance value over 50 iterations. \\

\noindent {\bf Performance of {$\boldsymbol{\iotdev}$}:}  \label{subsubsec:evaluation-iotdev}
\paisa overhead on \iotdev is measured in two phases: \boot and \runtime. 

\boot \ comprises the time taken for device initiation (\initdev), \timesync, and \announcement. 
During \initdev, \iotdev initiates the MCU itself and peripherals including timers, sensors, actuators, and network interfaces.
Next, during \timesync, \iotdev initiates its WiFi module in Station mode to connect to \mfr using UDP.
After a successful connection, \iotdev and \mfr communicate to synchronize the former's clock.
Then, \iotdev executes \announcement to issue its first \announcemessage.
As shown in Table \ref{table:eval-iotdev}, the time for \initdev is 9.66{\it ms} with 
negligible standard deviation. Whereas, average latency of \timesync is 1,076{\it ms} 
with a significant deviation of 187{\it ms}. This is because \timesync includes
network delay and all messages exchanged between the parties. 
Another reason for the high mean latency of \timesync is due to: (a) two signing operations during 
\syncreq and \syncack, and (b) one verification operation during \syncresp. Each ECDSA signing/verification 
operation takes $\approx230${\it ms} at {\it 150MHz}.
Finally, \announcement takes 236{\it ms}, which includes one signing operation and a beacon frame transmission.
Adding all these, the total boot time is about 1.3{\it s}, which is mostly due to \timesync and \announcement.
However, since this happens infrequently, we believe it is reasonable.

\begin{figure}[t]
  \centering
  \captionsetup{justification=centering}
  \includegraphics[width=1\columnwidth]{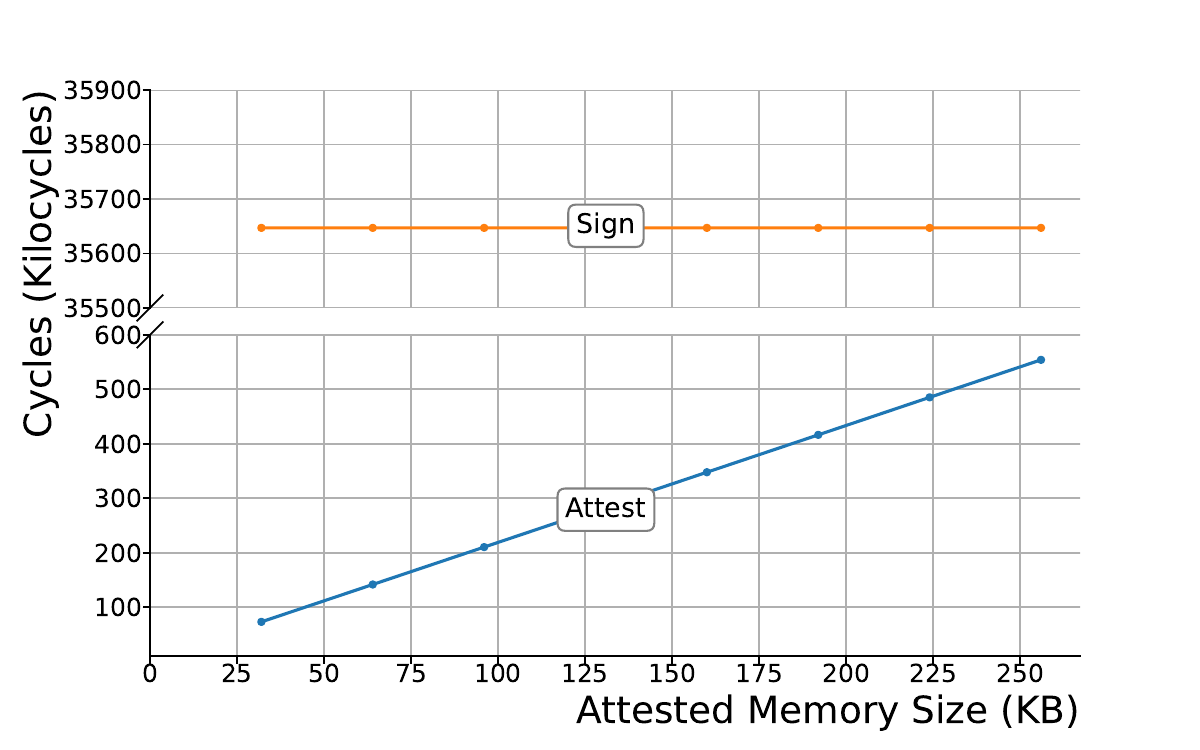}
  \vspace{-.3cm}
  \caption{\small \paisa \announcement Overhead on \iotdev at \runtime.} 
  \label{fig:eval-iotdev}
  \vspace{-.3cm}
\end{figure}

\runtime overhead stems from the \paisa \announcement module.
Figure \ref{fig:eval-iotdev} shows the performance of \announcement with variable size of the attested region. 
The latency for generating  and signing an \announcemessage is constant since the signature 
is over a fixed-sized value. Attestation latency grows linearly with the attested memory 
size since it requires hashing. However, signing takes significantly longer, about 230{\it ms}, 
than attestation, which only requires 1{\it ms} for 64KB. 
This is because public key operations naturally take more time than hashing.
Therefore, \announcement latency almost equals that of one signature operation.
Also, the software size of mid-to-low-tier devices is typically under 100KB.
Even if it reaches 1MB, attestation would take only $\approx~16${\it ms}, which is 14 times less than one signature.
Furthermore, during \announcement, the runtime overhead of the network interface is 
negligible, amounting to $\approx~135\mu$s, which has minimal impact on overall latency.

\begin{table}[h]
  \vspace{-.7em}
  \footnotesize
  \centering
  \resizebox{0.9\columnwidth}{!} {
     \begin{tabular}{|c||c|c|c|c|} \hline
     \multirow{2}{*}{\paisa Procedure}  & \multicolumn{2}{c|} {Cycles}  & \multicolumn{2}{c|} {Time @ 150MHz (ms)} 
     \\ \cline{2-5}
                                        &  Mean           & Standard Deviation  & Mean     & Standard Deviation
                                        \\ \Xhline{2\arrayrulewidth}
                        \initdev        & 1,449,461       & 121              & 9.66     & 0.02 \\ \hline
                        \timesync       & 161,386,850     & 28,129,473       & 1,075.91 & 187.53 \\ \hline
                        \announcement   & 35,431,478      & 87,119           & 236.21   & 0.58 
                        \\ \hline
     \end{tabular}
  }
  \caption{\small \paisa Overhead on \iotdev at \boot.} \label{table:eval-iotdev}
  \vspace{-.7cm}
\end{table}

\noindent {\bf Performance of {$\boldsymbol{\usrdev}$}:}\label{subsubsec:eval-usrdev}
The latency of \reception application is shown in Table \ref{table:eval-otherdev}.
It takes 1,070{\it ms} with a deviation of 247{\it ms} to receive one \announcemessage. 
This large deviation is due to two factors: the time to fetch \devmanifest depending on 
network delay and frequency, plus context switching time on the smartphone.
Note that Google Pixel 6 has heterogeneous cores (2 cores 
@ 2.8GHz, 2 cores @ 2.25GHz, and 4 cores @ 1.8GHz), thus, the overall frequency is represented as 
[1.8-2.8]GHz in Table \ref{table:eval-otherdev}.
Despite it taking 1{\it s} for one message, there is not much impact in case of multiple \iotdev-s, 
because \announcemessage processing can be done concurrently via threading (\verb+AsyncTask+).
Therefore, upon launching the Reception app, the delay in receiving most announcements is expected to be within a few seconds.

\begin{table}[h] 
  \vspace{-.7em}
  \footnotesize
  \centering
  \resizebox{0.9\columnwidth}{!} {
     \begin{tabular}{|c|c||c|c|} \hline
     \multirow{2}{*}{Device} & \multicolumn{1}{c||}{\paisa}  & \multicolumn{2}{c|} {Time (ms)} 
     \\ \cline{3-4}
                                 &  Procedure   &  Mean       & Standard Deviation
     \\ \Xhline{2\arrayrulewidth}
              \mfr @ 2.6GHz      &  \timesync       &  5.60       & 2.77          \\ \hline
              \usrdev @ [1.8-2.8]GHz   &  \reception      &  1070.34    & 247.00   
     \\ \hline
     \end{tabular}
  }
  \caption{\small \paisa Overhead on \mfr and \usrdev.} \label{table:eval-otherdev}
  \vspace{-3.0em}
\end{table}

\noindent {\bf Performance of {$\boldsymbol{\mfr}$}:} 
\timesync has one signing and two verification operations which take about 1{\it ms} each at 2.6GHz.
Hence, the average latency of \timesync is 5.6{\it ms} with a deviation of 2.77{\it ms}, mostly due to network delay. 
This latency is reasonable, despite \mfr handling multiple devices, because they can be served in 
parallel. Moreover, \timesync only occurs at reboot which is quite infrequent for each \iotdev. 

\noindent {\bf {$\boldsymbol{\devmanifest}$} size:} 
Many factors, such as device description, cryptographic algorithm, key size, type of certificates, 
and encoding method used in certificates, influence the size of \devmanifest. 
Thus, \devmanifest can vary from a few to a few hundred KB-s.
The size of \devmanifest used in our evaluation is 2,857 bytes. 

\noindent {\bf TCB size:} \label{subsubsec:evaluation-tcbsize}
As mentioned in Section \ref{subsec:impl_iotdev}, \paisa TCB consists of software in TZ-M of the
main NXP board and the driver in the network ESP32 board. 
On the main board, the TCB is 184KB (includes Mbed TLS), and 682KB on the network board (includes 
the network stack).

\section{Discussion \& Limitations}\label{sec:limitation}

We now discuss some limitations of \paisa and potential mitigations. \\

\noindent {\bf Run-time Overhead:} 
To measure run-time overhead on \iotdev, we define CPU utilization ($U$) as the percentage of CPU cycles that can be used by the normal application amidst the announcements, denoted by $U = \frac{t_{normal}}{t_{normal} + t_{ann}}$.
Here, $t_{normal}$ is the CPU cycles for the normal application between two announcements, which equals to \announceinterval, and $t_{ann}$ is the time taken for one announcement, which is nearly 250 {\it ms} (from Section \ref{subsubsec:evaluation-iotdev}).
So if $\announceinterval = 1s$, then $U = 80\%$ of normal utility, which is not good for general applications.
If $\announceinterval = 100s$, then $U = 99.7\%$, but it is not good for the users since they could not be aware of \iotdev up to 100s.
Therefore, depending on the application, there is a desired balance between the normal utility and the announcement interval.

There are other ways to reduce the overhead of \paisa. 
 
If the normal application binary size is large, \attestinterval can be increased to lower the 
overhead at every \announceinterval.
However, this might not yield much of a reduction since, as can be seen in Figure \ref{fig:eval-iotdev},
signing incurs higher overhead than attestation. Therefore, we consider the following option.

If the activity schedule of \iotdev is known, it can pre-compute multiple \announcemessage-s during 
idle time and later release one at a time. In this case, amortized (real-time) overhead would be significantly 
lower, since it would be only due to broadcasting \announcemessage.
For example, a smart speaker can precompute a day's worth of announcements at midnight and gradually release them. 
However, this approach is only applicable to devices that are not real-time and/or safety-critical. 
Also, in settings where a group of very low-end devices (e.g., smart bulbs) is connected to a 
local hub or controller, the latter can act as a \paisa proxy, i.e., it can broadcast a collective
announcement on behalf of the entire group of its constituent devices. 

\noindent {\bf Compatibility with other RoTs:}
\paisa can be applied to any architecture that offers a secure timer and a secure network interface. 
ARM TrustZone-A (TZ-A) is widely available in higher-end IoT devices that rely on ARM Cortex-A-based microprocessors 
(e.g., Raspberry Pi and Rock Pi). Since TZ-A offers similar guarantees to TZ-M, \paisa can be directly 
realized on the former.

For lowest-end MCUs, such as TI MSP430 \cite{TI-MSP430} and AVR ATMega8 \cite{atmel_specs}, an active RoT, called 
GAROTA\cite{garota}, offers a secure timer, GPIO, and UART peripheral support based on some additional custom hardware. 
\paisa can be applied to GAROTA by extending the secure timer TCB of GAROTA to include periodic announcements.

Furthermore, there is a software-based MultiZone TEE \cite{multizone} for RISC-V-based MCUs.
Relying on Physical Memory Protection Unit (PMP), Multizone divides memory and peripherals 
into well-isolated regions, called Zones, which are configured at compile-time.
\paisa can be implemented as one of the Zones with a timer peripheral and a network peripheral assigned to it.  

\noindent {\bf Compatibility with Other Network Interfaces:}
We believe that \paisa is compatible with other network interfaces besides WiFi, such as Bluetooth-Low-Energy 
and Cellular. For example, with Bluetooth version 5.0 and above, devices scan for other nearby devices by broadcasting 
packets that contain the sender address and advertising payload which can be up to 255 bytes.  
A \paisa announcement (116 bytes) can easily fit into this payload. 

\noindent {\bf Secure Update on {$\boldsymbol{\iotdev}$}:} To support secure software updates on \iotdev, \mfr or software 
vendors can initiate an update request by sending the new software along with its authorization token. This token is 
generated using a private key for which the corresponding public key is known to \iotdev. 
Implementing this process requires extending \paisa TCB to include token verification and update installation.
We expect that this update procedure can be implemented in a manner similar to existing frameworks,
such as \cite{scuba,uptane,casu}. 

\noindent 
{\bf User Linkage:}
There are both practical and conceptual techniques for anonymous retrieval that can be used to fetch \devmanifest-s.
The former include Tor, Mix Networks (e.g., Jondo and Nym), and peer-to-peer networks (e.g., I2P, Freenet). 
They all facilitate anonymous communication, however, their use might be illegal in some jurisdictions, 
while in others their use might be impractical due to additional requirements, such as Virtual Private Network (VPN).
Conceptual techniques include privacy-preserving cryptographic constructs, such as Private Information Retrieval (PIR) 
\cite{pir-tor,pir-query} and Oblivious RAM (ORAM) \cite{oram-cloud,oram-dp}. Using these types of techniques would 
require building customized ``wrappers'' for \paisa.

\noindent {\bf \paisa TCB:} 
As discussed in Section \ref{subsubsec:evaluation-tcbsize}, though the TCB size of the main device is small, the total size 
(including the network driver) increases the attack surface. 

Unfortunately, this is unavoidable because \paisa's main objective is guaranteed announcements which necessitates its
reliance on a trusted network interface. However, to alleviate this problem, we suggest pruning the network module to 
only contain what is absolutely necessary. 
For example, \paisa only requires the driver to establish a UDP connection with \mfr and broadcast WiFi beacon frames. 
The rest of the driver module (including TCP, HTTP, etc.) can be removed, thus significantly reducing the binary size. 
However, if normal applications want to use these protocols (via the secure stub mentioned earlier), 
the driver has to retain them.

\noindent {\bf Exclusive Network Module:} 
To ensure protection from DoS attacks, \paisa requires exclusive access to a network peripheral on \iotdev. 
This is because a shared network interface can be easily exploited by \sadv by keeping the interface 
busy and not allowing \announcemessage packets to be sent out. 

However, reserving a network interface exclusively for TCB use is expensive, since
the \iotdev budget might not permit an additional interface (in terms of cost and/or energy) for normal use.
To address this concern, we suggest using techniques such as \cite{rttee,minimuniotee,SeCloak} that involve 
a secure stub that shares peripherals between secure and non-secure programs. 
The main idea is to lock the network interface as a trusted peripheral controllable only by TZ-M. 
Also, a stub is implemented in the secure region that carefully parses inputs and relays them to the trusted interface.
This stub is made available to normal applications by exposing an NSC function callable from the normal region. 
Furthermore, the stub must also implement a scheduling queue for handling requests from both secure and non-secure 
applications. This way, there is no need to equip \iotdev with an additional interface.
We implement a basic functionality of this approach as a proof-of-concept. It is available as part of \cite{paisa-code}.
Nonetheless, we emphasize that, for the "timeliness" property of \paisa, the \announcement module is always given higher 
priority for accessing the network interface.

\noindent {\bf Role of \mfr:}
\paisa relies on \mfr for \timesync and hosting a database for \devmanifest. 
If the number of \iotdev-s provisioned by \mfr is high and \mfr is consistently overloaded with requests, 
we suggest using helper third-party servers in the local area of deployment.
Of course, such servers must be certified by \mfr to prove their authenticity when 
responding to \timesync and \devmanifest retrieval requests.

\section{Related Work}\label{sec:relatedwork}
Related work can be classified into six categories: 

\noindent {\bf Active RoTs} proactively monitor activity on MCUs to prevent (or minimize the extent of) compromises.
For example, \cite{garota,pfb,casu} are co-design (hardware/software) architectures that guarantee the execution of critical software even when all device software is compromised. 
\cite{pfb} guarantees sensor data privacy by letting only authorized software access sensor data via secure GPIO peripherals. 
On the other hand, \cite{casu} prevents code injection attacks by allowing only authorized software to run on the MCU while preventing any other software from modifying it except via secure authorized updates.
Whereas, \cite{awdt-dominance,lazarus-effect} rely on ARM TrustZone or a similar class of MCUs to protect devices from being "bricked", by resetting and updating the device whenever it does not respond to a watchdog timer. 

\noindent {\bf Remote Attestation:} 
There is a large body of research proposing remote attestation architectures on wide-range of devices. 
\cite{smart,sancus,vrasedp,simple,tytan,trustlite,TPM,seshadri2004swatt,apex,flicker,SCHELLEKENS200813,parsel} propose attestation architectures for MCUs. 
There are also other architectures such as \cite{litehax,cflat,lofat,tinycfa,atrium,oat,dialed,geden2019hardware} that discuss runtime attestation techniques, including control-flow, data-flow attestation,  for low-end MCUs.
All the aforementioned attestation architectures can be integrated with active RoTs mentioned earlier to enable \paisa.

For servers and high-end IoT, there are TEE architectures such as Intel SGX \cite{sgx}, AMD SEV \cite{sev}, Sanctum \cite{sanctum} and Keystone \cite{keystone} that provide attestation APIs for attesting in-enclave applications. 
However, these are not applicable for \paisa because \paisa attests and reports the normal region instead of the secure region. 

\noindent {\bf ARM TrustZone:} 
Lots of prior work leveraged TrustZone to improve the security of systems from various perspectives. 
\cite{tz-regulating,tz-sara,tz-dice} use TZ-A as an authorization tool for non-secure applications.
\cite{tz-regulating} proposes an authorization architecture to regulate smaller user devices connected to IoT hubs, enabled by TZ-A.
\cite{tz-sara} implements a user authentication scheme based on TZ-A on smartphones.
Besides these, TZ-M is also used to enhance security in several constrained settings, e.g., to optimize secure interrupt latencies \cite{tz-m-sbi}, improve real-time systems \cite{tz-m-rttee}, mitigate control-flow attacks \cite{cflat,tz-m-faslr}, and add support for virtualization \cite{tz-m-virtualize}.
Similarly, in \paisa, we use TZ-M to trigger announcements at regular intervals. 

\noindent {\bf Hidden IoT Device Detection:} 
To detect hidden IoT devices in unfamiliar environments, there are a few approaches proposed in recent years.
"spyware" solutions such as \cite{NLJD,bugdetector} are popular detectors; however, the detector should be in close proximity to the IoT device.
\cite{mmwave} designs specialized hardware -- a portable millimeter-wave probe -- to detect electronic devices. 
\cite{lapd} leverages the time-of-flight sensor on commodity smartphones to find hidden cameras.
However, they either take significant time or require specialized hardware to detect the devices.
Moreover, they can only detect IoT devices, but cannot identify them. 

On the other hand, \cite{lumos,snoopdog,devicemien2019,cameraspy} observe WiFi traffic to identify hidden devices. 
In particular, \cite{lumos} monitors coarse attributes in the WiFi 802.11 layer to classify IoT devices. 
\cite{snoopdog} establishes causality between WiFi traffic patterns to identify and localize an IoT device. 
\cite{devicemien2019} uses autoencoders for automatically learning features from IoT network traffic to classify them.
However, all the aforementioned techniques rely upon probabilistic models, hence, they can be error-prone, especially when there are newer devices or when the adversary is strong enough to bypass the detection logic; moreover, they are computationally intensive.
Conversely, \paisa takes a systematic approach to make users aware of the devices with minimal computation on their end.
Furthermore, \paisa announcements convey more information regarding the device such as its revocation status, software validity, and complete device description, which is not possible with other approaches. 

\noindent {\bf Broadcasting Beacon Frames:} 
\cite{beacon-stuffing} proposes a technique, Beacon-stuffing, that allows Wi-Fi stations to communicate with APs without associating with any network. 
Subsequently, many applications of Beacon-stuffing have been introduced over the past decade. 
\cite{bf2fa} uses beacon frames to figure out if a given device is physically located nearby a user device while the user is using the former for Two-Factor Authentication. 
\cite{wi-sl} achieves two-way data encryption transmission by injecting custom data into the probe request frame. 
\cite{WiFiHonk} proposes a smartphone-based Car2X communication system to alert users about imminent collisions by replacing the SSID field in the beacon frame with the alert message. 
Following the 802.11 standard, \cite{Gupta2012InformationEI} shows that custom information can be embedded in a beacon frame by modifying vendor-specific fields.

\noindent {\bf IoT Privacy:}
Some prior work focused on enhancing user privacy in the context of IoT via Privacy Assistants (PA-s) 
user notices, and consent. PA-s \cite{hong2004architecture,pa2,pa3} provide users 
with an automated platform to configure their privacy preferences on nearby IoT resources. For example, 
a recent study \cite{iotprivacy-ppa} interviews 17 participants to learn user perceptions of several 
existing PA-s and identifies issues with them. It then suggests ideas to improve PA-s in terms of automated consent, 
and helping them opt out of public data collections. \cite{iotprivacy-designspace} explores a comprehensive 
design space for privacy choices based on a user-centered analysis by organizing it around five dimensions 
(e.g. type, functionality, and timing). It also devises a concrete use case and demonstrates an IoT privacy 
choice platform in real-world systems.

Furthermore, some research efforts have explored privacy and security labels (akin to food nutrition labels) for IoT devices. 
For example, \cite{iotprivacy-label} suggests a set of IoT privacy and security labels based on interviews and surveys. 
It identifies $47$ crucial factors and proposes a layered label approach to convey them. \cite{iotprivacy-attributes} 
conducts a survey with $1,371$ online participants to evaluate the privacy factors proposed in prior research with two key dimensions: 
an ability to convey risk to consumers and an impact on their willingness to purchase an IoT device. 
Also, the study yields actionable insights on optimizing existing privacy and security attributes of IoT labels. 
Similarly, \cite{iotprivacy-consumers} conducts a survey with $180$ online participants in order to evaluate the impact
of five security and privacy factors (e.g. access control) on participants' purchase behaviors when 
individually or collectively presented with an IoT label. The study underscores participants' willingness to 
pay a substantial premium for devices with better security and privacy practices.

These prior results are valuable and relevant to this paper since they provide guidelines for which privacy-related 
factors should be reflected in \devmanifest and how to utilize them in order to attain acceptable user experience with effective privacy configurations.

\section{Conclusions}\label{sec:conclusion}
This paper suggests taking a systematic approach to making IoT devices \privacyaware \ by advocating that 
devices periodically inform nearby users about their presence and activity. As a concrete example of this approach,
we presented the design and construction of \paisa: a secure and \privacyaware\ TEE-based architecture that 
guarantees secure periodic announcements of device presence via secure timer and network peripherals. 
We implemented \paisa as an end-to-end open-source prototype \cite{paisa-code} on: (1) an ARM Cortex-M33 device equipped with 
TrustZone-M that broadcasts announcements using IEEE 802.11 WiFi beacons, and (2) an Android-based app
that captures and processes them. The evaluation shows that \iotdev takes $236$ms 
to transmit an announcement and it only takes $1$sec for the app to process it. 

\noindent{\bf Acknowledgements}: We thank ACM CCS’23 reviewers for
constructive feedback. This work was supported in part by funding from NSF Awards
SATC-1956393, SATC-2245531, and CICI-1840197, NSA Awards
H98230-20-1-0345 and H98230-22-1-0308, as well as a subcontract from Peraton Labs.

\balance
\bibliographystyle{abbrv}
\bibliography{references}

\end{document}